\documentclass[twocolumn,showpacs]{revtex4}
\usepackage[xdvi]{graphicx}
\usepackage{dcolumn}
\usepackage{amsmath}

\newcommand{\bra}{\langle}
\newcommand{\ket}{\rangle}

\newcommand{\bv}[1]{{\boldsymbol #1}}

\begin{document}

\title{Thermodynamic formula for the cumulant generating function of 
time-averaged current}

\author{Takahiro Nemoto and Shin-ichi Sasa}
\affiliation{Department of pure and applied Sciences,
The University of Tokyo, Tokyo, 153-8902, Japan}
\date{\today}

\begin{abstract}
The cumulant generating function of time-averaged current is studied from
an operational viewpoint. Specifically, for
interacting Brownian particles under
non-equilibrium conditions, we show that the first derivative of 
the cumulant generating function is
equal to the expectation value of the current in a modified system
with an extra force added, where the modified system is 
characterized by a variational principle.
The formula reminds us of
Einstein's fluctuation theory
in equilibrium
statistical mechanics. Furthermore,
since the formula leads to the fluctuation-dissipation relation when the
linear response regime is focused on, it is regarded as an
extension of the linear response theory to that valid beyond the
linear response regime.
The formula is also related to previously known theories 
such as the Donsker-Varadhan
theory, the additivity principle, and the least dissipation principle, 
but it is not derived from them.
Examples of its application are
presented for a driven Brownian particle on a ring subject to a periodic
potential.
\end{abstract}

\pacs{05.40.-a, 05.70.Ln, 02.50.Ey}

\maketitle


\section{Introduction}


Entropy $S$ as a function of extensive variables 
characterizes macroscopic properties of materials in a unified 
manner \cite{callen}. Its functional form for a given material is completely determined by thermodynamic measurements 
of the heat capacity, the compressibility and the other relevant susceptibilities.
Let $X$ be 
an unconstrained variable of a composite isolated system. 
We denote the $X$ dependence of $S$ by $S(X)$. Then,  
the equilibrium value of $X$ is determined by 
\begin{equation}
X_*=\underset{X}{\rm argmax}[S(X)],
\label{ext}
\end{equation}
which is an example of the variational principle of 
thermodynamics. Furthermore, 
according to Einstein's fluctuation theory \cite{callen}, 
the probability density of $X$ is expressed as 
\begin{equation}
{\rm Prob}(X) \simeq e^{ \left ( S (X )-S(X_*) \right )/k_{\rm B}},
\label{ein}
\end{equation}
where $k_{\rm B}$ is the Boltzmann constant,
and  $\simeq $ represents the asymptotic equivalence 
of the exponential factor in the thermodynamic limit.
More precisely, let $N$ be the number of 
the degrees of freedom. Then,
the scaled entropy $s(x)=\lim_{N\rightarrow\infty}S(X)/N$
with $x=X/N$ fixed satisfies
\begin{equation}
\frac{1}{k_{\rm B}}\left ( s(x)-s(x_*) \right )
=\lim_{N \to \infty} \frac{1}{N} \log {\rm Prob}(x).
\label{rate}
\end{equation}
In the probability theory, the right-hand side provides
a large deviation function (up to the minus sign) \cite{Dembo_Zeitouni}.
That is, the scaled entropy $s(x)$ is  equivalent  to 
the large deviation function. 


Now, suppose that the scaled entropy $s(x)$ is
defined by (\ref{rate}) with the principle of the equal weight
for microscopic states.
Then, the variational principle
(\ref{ext}) can be derived from the definition. A surprise
in this approach is that the entropy thus defined characterizes
thermodynamic properties completely through the fundamental 
relation of thermodynamics. After establishing the connection
to thermodynamics, one can obtain the scaled entropy
by thermodynamic measurements.
In other words, the large deviation function,
which characterizes macroscopic fluctuations, is obtained
without measuring fluctuations.  Such formulas are not 
restricted to the case of fluctuations in isolated 
systems.  As demonstrated for a simple model in Appendix \ref{appendix_thermodynamics},
several formulas are derived. One example is that 
a scaled cumulant generating function of a thermodynamic variable, 
which is given by the Legendre transform of the large deviation 
function, is related to a free energy density in thermodynamics.
From this relation, it is found that the derivative of the scaled 
cumulant generating function with respect to its argument 
is equal to the expectation value of the thermodynamic variable
in a modified system with an extra external force 
added. See (\ref{G_equilibrium}) for such an example.
The relations between large deviation functions and thermodynamic
functions are thought to be well-known, and the relations lead
to fluctuation-response relations, one of which, 
connects the susceptibility and the intensity of fluctuations
of magnetization \cite{callen}.


The main difference of non-equilibrium systems from equilibrium systems
is the existence of a current. In particular, 
the expectation value of the current in a linear response regime
is connected with the intensity of fluctuations of time-averaged current
in equilibrium \cite{statphys2,Zwanzig}. This is one of fluctuation-dissipation
relations, which might correspond to fluctuation-response relations
in equilibrium statistical mechanics.
Since the time-averaged 
current appears to be important at least for small fluctuations 
in the linear response regime, it is natural to consider the large 
deviation function of time-averaged current.
Then, from
the analogy with thermodynamics, one may expect that there is a
formula that connects the large deviation function with a 
macroscopic quantity and that 
the fluctuation-dissipation  relations are derived from the formula
when the linear response regime
is focused on.
Indeed, Onsager and 
Onsager-Machlup 
attempted to express the large deviation function of time-averaged
current
in terms of dissipation functions 
\cite{Onsager, OnsagerII,Onsager_Machlup}, as  pointed out by Oono \cite{oono}.
This is successful for small fluctuations in the linear response 
regime, but the expression cannot be applied to general cases with 
large fluctuations.
 
 
In the last two decades, the  understanding of the large 
deviation function of time-averaged current has been
substantially developed. One remarkable result is the 
discovery of the fluctuation theorem which 
represents a simple and beautiful symmetry property of 
the large deviation function of time-averaged entropy 
production \cite{FT}.
The symmetry property was understood from the 
microscopic reversibility of the system and it was generalized
to identities for any quantities \cite{Gallavotti_cohen, Kurchan_fluctuation, Maes_fluctuation, Crooks_fluctuation, Lebowitz_Spohn}.
This progress enables us to revisit a linear response theory
and to re-derive known formulas including
a non-linear response formula 
\cite{KG} and 
an expression of steady state distribution \cite{Zubarev, McLennan}.
See  Ref. \cite{Hayashi-Sasa2} as a review of these re-derivations.
In addition to the re-derivation of the known formulas, the symmetry property
plays an essential role in deriving an extension of the
Clausius relation \cite{KNST, KNST_regular, Saito_Tasaki}
on the basis of a neat expression of steady state distribution \cite{Komatsu_Nakagawa}. 
As another direction of the progress,  the large deviation 
function of time-averaged current was explicitly calculated 
for lattice models \cite{Derrida_Lebowitz, Derrida_exactreview}. 
The results evolved into a 
variational form, which was called an additivity principle 
\cite{BD, BD2, BD3, Derrida_review}.  
When the system behavior is
assumed to be described by fluctuating hydrodynamics, the 
condition for the validity of the additivity principle can be 
derived from the analysis of the path-integral 
expression \cite{phase-transition1, BDGJL,
BDGJL2, phase-transition2, phase-transition4}. Note that 
the additivity principle is less universal than the fluctuation
theorem, but still characterizes a universal aspect of 
the large deviation functions of time-averaged current 
in a wide range class of systems. See Ref. \cite{Hurtado}
for a recent related study.


Based on these developments, we consider a possibility that
the large deviation function of time-averaged current is 
expressed by a time-averaged quantity as if the framework corresponds to Einstein's
fluctuation theory.
This formula,
if it exists, is regarded as an extension of Onsager's theory 
to that valid even for large fluctuations. 
Although the motivation seems natural, the problem
would not be solved easily. Nevertheless, since a new operational variational 
principle associated with the large deviation function 
of time-averaged current has been
found recently \cite{Nemoto_Sasa}, it is a good opportunity to study the problem.


In this paper, specifically, we study interacting Brownian 
particles under non-equilibrium conditions. The motion of the
particles are assumed to be described by a set of Langevin equations.
We then consider the large deviation function of  time-averaged 
velocity of the particles. The main results in this paper are 
formulas (\ref{cumulantvariational_expression}) and
(\ref{mainclaim_of_the_paper}). The formula (\ref{cumulantvariational_expression})
indicates that the scaled cumulant 
generating function, which is given by the Legendre transform 
of the large deviation function, is determined  from  a variational 
function in a modified system with an inhomogeneous external force added.
Essentially the same formula was reported in Ref. \cite{Nemoto_Sasa}.
The formula (\ref{mainclaim_of_the_paper})
claims that the derivative 
of the scaled cumulant generating function is equal to 
the time-averaged velocity in the modified system, which corresponds to 
(\ref{G_equilibrium}) in equilibrium statistical mechanics.
The formula (\ref{mainclaim_of_the_paper}) also turns out to be 
an extension of the fluctuation-dissipation relation.
Since these new formulas take simple forms, we expect that our results
advance the understanding of large deviation function of time-averaged
quantity.


This paper is organized as follows. 
In section \ref{section_model}, we introduce a model and define
quantities we study in this paper. In section \ref{section_main},
we first summarize main results of the paper and after that we derive
them.
In section \ref{Section_whatismaximized}, 
we address three remarks on the variational function
that appears in our main results. Concretely, 
we re-express the variational function in terms of entropy 
production rates, mention a similarity with the Donsker-Varadhan 
formula, and explain a connection to the principle of the least dissipation
of energy. In section \ref{Section_application},
we present a few applications of our formula by restricting our study to
a single Brownian particle on a ring.
First, on the basis of our formula, we re-derive exact formulas of the  diffusion 
constant and the mobility in a compact manner.
Second, starting from our formula, we show a variational expression
associated with an additivity principle. Third,
we propose a method for experimentally obtaining the scaled cumulant generating 
function.
The final section is devoted to concluding remarks.
Some technical details and related subjects are discussed
in Appendices. In the argument below, the Boltzmann constant
$k_{\rm B}$ is set to unity.


\section{Model}
\label{section_model}



We consider $N$ colloidal particles immersed in a solvent at temperature $T$.
We describe the system by a set of positions of the particles in a
$d$-dimensional region $\Omega$.
For mathematical simplicity, we impose periodic boundary conditions. 
As a typical example, one may consider a three-dimensional
region, where one direction is confined by walls and no walls
in other directions. Even such a case is
included in our model by setting up a periodic potential
whose height is extremely large at the walls.
Another example of the region $\Omega$ is
a one-dimensional circuit which can be generated by
an optical potential.


We denote the positions of 
all the particles by $\bv x \equiv (x_i)_{i=1}^{Nd}$, where $x_{(k-1)d+l}$
represents the $l$-th component of the position of the $k$-th particle.
In this paper, bold-face fonts (e.g. $\bv x$) represent
$Nd$-dimensional vectors. 
We also express the position of the $k$-th particle by 
the  $d$-dimensional vector $\vec{r}_k = (x_{(k-1)d+1}, x_{(k-1)d+2},..., x_{kd})$.


The force acting on the particles consists of three types of force, 
(i) an external force, (ii) an interaction force exerted from 
other particles, and (iii) a force from solvent molecules. 
In (i), the external force is further divided into
a homogeneous part $\bv f = (f_i)_{i=1}^{Nd}$
and a periodic part determined by a potential
$\sum_{k=1}^{N}U_k(\vec{r}_k)$.
Since the existence of $\bv f$ makes the system out of
equilibrium, we call $\bv f$ a driving force.
In (ii), the interaction force is determined by
a two-body interaction potential $U_{k,l}(\vec{r}_k-\vec{r}_l)$ 
between the $k$-th particle 
and the $l$-th particle under the assumption that
three- and higher-body interactions among particles are not taken into account. 
The total potential $U(\bv x)$ is written as
\begin{equation}
U(\bv x) = \sum_{k=1}^N U_k(\vec{r}_k) 
+ \frac{1}{2}\sum_{k=1}^{N}\sum_{l\neq k} U_{k,l}(\vec{r}_k-\vec{r}_l) .
\end{equation}
The total force 
$\bv F (\bv x)$ that originates from (i) and (ii) is expressed as
\begin{equation}
\bv F (\bv x) = \bv f - \frac{\partial}{\partial \bv x} U(\bv x).
\end{equation}
In (iii), the force from solvent molecules is divided into two parts,
a friction force and a fluctuating force, where we assume that
hydrodynamic interactions are ignored. 
The friction force is proportional to the particle velocity
$\dot x_i = dx_i/dt$ so that it can be written in the form of
$-\dot x_i  \gamma_i$ with friction constants $( \gamma _i )_{i=1}^{Nd}$.
We assume that the fluctuating force is described as Gaussian white noise
satisfying the fluctuation-dissipation relation of the second kind. 
That is, it is expressed
by $\sqrt{2T\gamma_i}\xi_i(t)$, 
where $\xi_i(t)$ satisfies
\begin{equation}
\langle \xi _i(t) \rangle =0, \label{gaussian_noise1}
\end{equation}
\begin{equation}
\langle \xi _i(t) \xi _j(s)  \rangle = \delta _{i,j} \delta(t-s).
\label{gaussian_noise2}
\end{equation}


From these arguments, we obtain the equations of motion
for colloidal particles.
By neglecting inertial effects, these are written as
\begin{equation}
\dot x_i = \frac{1}{\gamma _i}F_i({\bv x}) + \sqrt{\frac{2T}{\gamma _i}}  
\xi _i, 
\label{langevin}
\end{equation}
where $i=1,2,\cdots,Nd$. The equations are called Langevin equations \cite{Gardiner} 
and have been studied in many situations including
Brownian motors \cite{Review_Reimann}, polymer dynamics \cite{Doi_Edwards}
and dynamics of mesoscale objects in solids. 
See Introduction of Refs. \cite{Reimann, Reimann2} and Chap. 11 of Ref. \cite{Risken}.
It should be noted that some recent discoveries of non-equilibrium relations
have been tested in experiments corresponding to (\ref{langevin})
with $N=d=1$ \cite{Bechinger,Ciliberto,Toyabe}.


We assume that the particles obey a probability density $P_0(\bv x)$ 
at $t=0$. 
After that ($t>0$), the particles move according to (\ref{langevin}). 
We denote by $\langle f \rangle$ the expectation value of $f$ with respect
to the initial distribution $P_0(\bv x)$ and history of $\xi_i(t)$.
We define the probability density of $\bv x(t)=\bv x$ by $P^{\bv F}(\bv x , t) \equiv \left \langle 
\delta \left (\bv x(t)-\bv x \right ) \right \rangle.$
Here, the superscript ${\bv F}$ of $P^{{\bv F}}(\bv x , t)$
represents $\bv F(\bv x)$ in (\ref{langevin}).
We use this superscript throughout the paper, 
because the force dependence plays an important role 
in our formulation.
The evolution equation for $P^{\bv F}(\bv x , t)$ is derived as
\begin{equation}
\frac{\partial P^{\bv F}(\bv x,t)}{\partial t} 
= -\sum_{i=1}^{Nd} \frac{\partial j^{\bv F}_i(\bv x,t)}{\partial x_i}
\label{model_continuity}
\end{equation}
with a probability current
\begin{equation}
j^{\bv F}_i(\bv x,t) 
= \frac{P^{\bv F}(\bv x,t)}{\gamma_i}F_i(\bv x)\label{model_current}
- \frac{T}{\gamma_i} \frac{\partial P^{\bv F}(\bv x,t)}{\partial x_i} .
\end{equation}
The equation is called the Fokker-Planck equation \cite{Gardiner}.


The stationary distribution and current, which  are 
denoted by $P^{\bv F}_{\rm st}(\bv x)$ and $J^{\bv F}_{{\rm st},i}(\bv x)$, 
respectively, are determined by
\begin{equation}
J_{{\rm st},i}^{{\bv F}}({\bv x}) 
= \frac{P_{\rm st} ^{{\bv F}}(\bv x)}{\gamma_i} F_i({\bv x})
 -\frac{T}{\gamma_i}
\frac{\partial P_{\rm st} ^{{\bv F}} (\bv x)}{\partial x_i} , 
\label{stationarryrho}
\end{equation}
\begin{equation}
\sum_{j=1}^{Nd}\frac{\partial }
{\partial x_j} { J^{\bv F}_{{\rm st},j}}({\bv x}) = 0.
\label{stationarrycontinue}
\end{equation}
The expectation value of 
a quantity $A\left [(\bv x(s))_{s=0}^t \right ]$ 
in the steady state is denoted by $\left \langle A\right \rangle _{\rm st}^{\bv F}$.
Mathematically, the expectation value is calculated
by replacing initial distribution $P_0(\bv{x})$ by $P_{\rm st}^{\bv F}(\bv{x})$.


In this paper, we particularly focus on  statistical 
properties  of time-averaged velocities
\begin{equation}
\bv V (\tau) \equiv \frac{1}{\tau} \int _{0}^{\tau}dt \dot {\bv x}(t).  
\label{average_velocity}
\end{equation}
The scaled cumulant generating function $G^{\bv F}({\bv h})$ 
of the time-averaged velocities is defined by 
\begin{equation}
G^{\bv F}({\bv h}) 
\equiv \lim_{\tau \rightarrow \infty} 
\frac{1}{\tau} \log \langle  e^{\tau \sum_{i=1}^{Nd} h_i
V_i(\tau)}   \rangle , 
\label{cumulantgeneratingfunction}
\end{equation}
where ${\bv h}=( h_{i})_{i=1}^{Nd}$. 
In the expansion form
\begin{equation}
G^{\bv F}(\bv h) = \sum_{k_1=0}^\infty\sum_{k_2=0}^\infty
\cdots  \sum_{k_{Nd}=0} ^\infty C^{\bv F}_{k_1,k_2,...,k_{Nd}} 
\frac{h_1^{k_1} h_2^
{k_2}...h_{Nd}^{k_{Nd}}}{k_1!k_2!...k_{Nd}!},
\label{cumulantexpansion}
\end{equation}
the coefficient $C^{\bv F}_{k_1,k_2,...,k_{Nd}}$ is related to
the cumulant  
$\langle V_{1}^{k_1}(\tau) V_{2}^{k_2}(\tau)...V_{Nd}^{k_{Nd}}
(\tau)\rangle_{\rm c}$ \cite{Gardiner}. Concretely, it can be shown that 
\begin{equation}
C^{\bv F}_{k_1,k_2,...,k_{Nd}}=\lim _{\tau \rightarrow \infty}
\langle V_{1}^{k_1}(\tau)V_{2}^{k_2}(\tau)...
V_{Nd}^{k_{Nd}}(\tau)\rangle_{\rm c}\tau ^{k-1}.
\label{C}
\end{equation}
Here, by substituting (\ref{langevin}) into (\ref{average_velocity})
and  taking the expectation value, we obtain
\begin{equation}
\left \langle V_i(\tau) \right \rangle 
= \int _{\Omega ^N} d\bv x \frac{1}{\gamma_i} F_i(\bv x) 
\left ( \frac{1}{\tau} \int_{0}^{\tau} dt  P^{\bv F}(\bv x , t) \right ) .
\label{stationary2}
\end{equation}
This leads to 
\begin{equation}
\left \langle V_i \right \rangle_{\rm st}^{\bv F} 
= \int _{\Omega ^N} d\bv x \frac{1}{\gamma_i} F_i(\bv x) 
P^{\bv F}_{\rm st}(\bv x ). 
\label{stationary3}
\end{equation}
Furthermore, 
by substituting (\ref{stationarryrho}) into (\ref{stationary3})
and by eliminating boundary terms due to periodic boundary conditions, 
we obtain
\begin{equation}
\left \langle V_i \right \rangle^{\bv F }_{\rm st} 
= \int _{\Omega ^N} d\bv x J_{{\rm st},i}^{\bv F }(\bv x) . 
\label{averaged}
\end{equation}


\section{Main results}
\label{section_main}


The main purpose of this paper is to find an expression
of $G^{\bv F}(\bv h)$ 
which shares a common structure in equilibrium  statistical 
mechanics. As reviewed in Appendix \ref{appendix_thermodynamics},
the cumulant generating function of a thermodynamic variable
is related to a free energy function and this relation 
leads to the important result that the first derivative of 
the cumulant generating function is 
equal to the expectation value of the thermodynamic 
variable in a modified system. See  (\ref{G_equilibrium}) for such an
example.
We want to have  a similar expression for $G^{\bv F}(\bv h)$.


Since the relation containing (\ref{G_equilibrium}) leads to the fluctuation-response
relation for thermodynamic variables, we expect that there might be a formula
similar to (\ref{G_equilibrium}), which leads to fluctuation-dissipation relations for currents.
In order to investigate this possibility, we focus on the linear response regime.
First, by noting $\left ( \partial G^{\bv F}(\bv h)/\partial h_i \right ) |_{\bv h =0} = 
\left \langle V_i \right \rangle_{\rm st}^{\bv F}$, we have
\begin{equation}
\frac{\partial G^{\bv F}(\bv h)}{\partial h_i} 
= \left \langle V_i \right \rangle _{\rm st}^{\bv F}
+\sum_{j=1}^{Nd}  h_j  
\left ( \frac{\partial^2 G^{\bv F}(\bv h)}{\partial h_i \partial h_j } \right )
{\Big |}_{\bv h=0}+ \mathcal{O}(\bv h^2).
\label{linear_response2}
\end{equation}
Here, the fluctuation-dissipation relation is written as 
\begin{equation}
\left \langle V_i \right \rangle_{\rm st}^{\bv F} 
= \frac{1}{2T} \sum_{j=1}^{Nd}
f_{j} \left ( \frac{\partial ^2G^{\bv F}(\bv h)}{\partial h_i \partial h_j} 
\right )
{\Big |}_{\bv h =0,\bv f=0}  + \mathcal{O}\left (\bv f ^2 \right ). 
\label{linear_response}
\end{equation}
See Refs. \cite{Agarwal,Risken} for a derivation. See also Refs. \cite{Lebowitz_Spohn, Hayashi-Sasa2} for a recent understanding of the
derivation on the basis of
the fluctuation theorem.
The substitution of (\ref{linear_response}) into the first term 
of the right-hand side of (\ref{linear_response2}) and the evaluation 
of the second term at $\bv  f=0$ yield
\begin{equation}
\begin{split}
\frac{\partial G^{\bv F} (\bv h)}{\partial h_i} 
=&\frac{1}{2T}
\sum_{j=1}^{Nd} \left (f_j + 2T h_j  \right ) 
\left ( \frac{\partial^2 G^{\bv F}(\bv h)}{\partial h_i \partial h_j } \right ) 
\Big{|}_{\bv h=0,\bv f=0}\\
&+ \mathcal{O}(\bv f^2 , \bv h^2,\bv f \bv h) .
\label{linear_response3}
\end{split}
\end{equation}
By using (\ref{linear_response}) again in
the right-hand side of (\ref{linear_response3}),
we obtain
\begin{equation}
\frac{\partial G^{\bv F}(\bv h)}{\partial h_i} 
= \left \langle V_i \right \rangle _{\rm st}^{\bv F + 2T\bv h}
+ \mathcal{O}(\bv h^2,\bv f ^2, \bv f \bv h).  
\label{differential_linear}
\end{equation}
This implies that when $\bv f$ and $\bv h$
are small, the first derivative of $G^{\bv F}(\bv h)$
is equal to the expectation value of the time-averaged velocity
in a modified system with an extra force $2T\bv h$ added.
Since (\ref{linear_response}) is derived from (\ref{linear_response2})
and (\ref{differential_linear}), the expression (\ref{differential_linear}) is equivalent to
the fluctuation-dissipation relation (\ref{linear_response}).


Now, we seek for an extension of (\ref{differential_linear}) to that
valid for any $\bv h$ and $\bv f$. 
Here, one may recall the fluctuation theorem as an extension of the fluctuation-dissipation relation, which claims
\begin{equation}
G^{\bv F}(\bv h) = G^{\bv F}(-\bv h-\bv f/T)
\label{fluctuation_teorem}
\end{equation}
in the system under consideration \cite{Lebowitz_Spohn}.
Although (\ref{linear_response}) is derived from (\ref{fluctuation_teorem}),
a clear extension of (\ref{differential_linear}) is not obtained as far as we attempted. Putting aside the fluctuation theorem, we consider the
problem from a different direction.
The basic idea is to consider a modified system with a spatially inhomogeneous 
extra force $\bv w(\bv x)$ added.
Precisely writing, the modified system is
described by the Langevin equation (\ref{langevin}) 
with the replacement of $\bv F(\bv x)$ by $\bv F (\bv x) + \bv w (\bv x)$.
We then define a functional 
$\Phi ^{\bv F}_{\bv h}( \bv w)$ by
\begin{equation}
\Phi^{\bv F}_{\bv h} (\bv w)
\equiv \sum_{i=1}^{Nd}
\left [ h_i \left \langle  V_i\right \rangle^{{{\bv F}+\bv w }}_{\rm st}
- \frac{1} {4T\gamma_i} \left \langle  w_i^2  
\right \rangle^{\bv F +  \bv w} _{\rm st}
\right ]  
\label{hannkannsuu}.
\end{equation}
The first result of this paper is a variational expression
\begin{equation}
G^{\bv F}(\bv h) =\max_{\bv w} \Phi _{\bv h}^{\bv F}( \bv w).
\label{cumulantvariational_expression}
\end{equation}
Let $ \bv w^{\bv F,\rm opt}_{\bv h}$ be the optimal $\bv w$ that 
maximizes (\ref{hannkannsuu}) for given $\bv h$ and $\bv F$.
That is,
\begin{equation}
\bv w^{\bv F, \rm opt}_{\bv h} = \underset{\bv w}{\rm argmax} 
\left [ \Phi^{\bv F}_{\bv h}
( \bv w) \right ].
\label{argment}
\end{equation}
The second result is a formula
\begin{equation}
\frac{\partial G^{\bv F}({\bv h})}{\partial h_i} 
= \left \langle  V_i\right \rangle^{{{\bv F}+
\bv w^{\bv F, \rm opt}_{\bv h} }}_{\rm st}.
\label{mainclaim_of_the_paper}
\end{equation}
This is thought to be the generalization of (\ref{differential_linear}). 
Indeed, there exists a positive function 
$\psi _{\bv h} ({\bv x})$ such that 
\begin{equation}
\bv w^{\bv F,\rm opt}_{\bv h } (\bv x)  
=2T\bv h + 2T\frac{\partial}{\partial \bv x}\log \psi _{\bv h} ({\bv x})
\label{wopt_psi}
\end{equation}
with a condition
\begin{equation}
\lim _{\bv h\rightarrow 0} \frac{\partial }{\partial x_i} 
\log \psi_{\bv h}(\bv x) =0.
\label{derivation_linear_response1}
\end{equation}

Before deriving the main results,
we confirm that the fluctuation-dissipation
relation (\ref{linear_response}) is obtained from our result (\ref{mainclaim_of_the_paper})
with (\ref{wopt_psi}). We start with
\begin{equation}
\frac{\partial G^{\bv F}({\bv h})}{\partial h_i} 
= \left \langle  V_i\right \rangle^{{{\bv f}-\frac{\partial U}{\partial \bv x}+
2T\bv h + 2T\frac{\partial \log \psi _{\bv h}}{\partial \bv x} 
}}_{\rm st}.
\label{derivation_start}
\end{equation}
From a basic property that there is no persistent current in systems
without a driving force, we have an identity
\begin{equation}
\left \langle V_i \right \rangle _{\rm st}^{ -
\frac{\partial U }{\partial \bv x} 
+2T\frac{\partial \log \psi_{\bv h}}{\partial \bv x} } 
=0.
\end{equation}
This leads to
\begin{equation}
\begin{split}
&\left \langle V_i \right \rangle _{\rm st}^{\bv f+2T\bv h -
\frac{\partial U }{\partial \bv x} 
+2T\frac{\partial \log \psi_{\bv h}}{\partial \bv x} } \\
&= \sum_{j=1}^{Nd} 
\lim _{\bv f^\prime \rightarrow 0} 
\left [\frac{ \partial \left \langle V_i 
\right \rangle _{\rm st}^{\bv f^\prime - \frac{\partial U}{\partial \bv x}
+2T\frac{\partial \log \psi_{\bv h}}{\partial \bv x}  } }
{\partial f_j^\prime}\right ](f_j+2Th_j) \\
&\quad + \mathcal O ((\bv f+2T\bv h)^2).
\label{derivation_linear_response2}
\end{split}
\end{equation}
By using (\ref{derivation_linear_response1}), we further rewrite
the right-hand side of (\ref{derivation_linear_response2}) as
\begin{equation}
\sum_{j=1}^{Nd} 
\lim _{\bv f^\prime \rightarrow 0} 
\left [
\frac{ 
\partial 
\left \langle 
V_i \right 
\rangle _{\rm st}^{\bv f^\prime - \frac{\partial U}{\partial \bv x}  } }
{\partial f_j^\prime}\right ](f_j+2Th_j) 
+ \mathcal O (\bv f^2,\bv h^2,\bv f \bv h).
\label{derivation_linear_response3}
\end{equation}
Thus, we arrive at
\begin{equation}
\begin{split}
\frac{\partial G^{\bv F}(\bv h)}{\partial h_i} =& 
\sum_{j=1}^{Nd} 
\lim _{\bv f^\prime \rightarrow 0} 
\left [\frac{ \partial \left \langle V_i 
\right \rangle _{\rm st}^{\bv f^\prime - \frac{\partial U}{\partial \bv x}} }
{\partial f_j}\right ](f_j+2Th_j)  \\
&+ \mathcal O (\bv f^2,\bv h^2, \bv f \bv h).
\label{linear_derivation}
\end{split}
\end{equation}
By differentiating (\ref{linear_derivation}) with respect to $h_j$ 
and setting $\bv f=0$ and $\bv h =0$, we obtain
\begin{equation}
\frac{\partial^2 G^{\bv F}(\bv h)}{\partial h_i \partial h_j}
\Big{|}_{\bv f=0, \bv h=0} = 2T
\lim _{\bv f^\prime \rightarrow 0} 
\left [\frac{ \partial \left \langle V_i 
\right \rangle _{\rm st}^{\bv f^\prime - \frac{\partial U}{\partial \bv x}} }
{\partial f_j}\right ].
\end{equation} 
This is equivalent to (\ref{linear_response}).


\subsection{Derivation}\label{Subsection_derivation}


The set of displacements of particles during the time interval $[0,t]$
is written as 
\begin{equation}
{\bv X}(t)=\int _{0}^{t}dt^{\prime}\dot{\bv x}(t^{\prime}),
\label{total_displacement}
\end{equation}
where it should be noted that
\begin{equation}
\bv X(t)=\bv V(t) t.
\label{XandV}
\end{equation}
By differentiating (\ref{total_displacement}) with respect to 
$t$, we obtain
\begin{equation}
\frac{dX_i}{dt} 
= \frac{1}{\gamma _i}F_i({\bv x}) + \sqrt{\frac{2T}{\gamma _i}} \xi_i
\label{langevin_X}.
\end{equation}
We denote the joint probability density of ${\bv x}(t)$ and 
$\bv X(t)$ by $p({\bv x},{\bv X},t)$. 
The idea of considering such a joint probability density of 
$\bv x$ and $\bv X$ (or accumulated work, heat) was employed in 
Refs. \cite{Speck_Seifert, Imparato_Peliti1, Imparato_Peliti2}.
From the Langevin equations (\ref{langevin}) and (\ref{langevin_X}), 
we derive the Fokker-Planck equation
\begin{equation}
\frac{\partial p}{\partial t} 
= \mathcal{L}_{\rm FP}^{({\bv x},{\bv X})}\cdot p ,
\label{FPequation}
\end{equation}
where the Fokker-Planck operator 
$\mathcal L_{\rm FP}^{(\bv x,\bv X)}$ is defined by
\begin{equation}
\begin{split}
&\mathcal{L}_{\rm FP}^{({\bv x},{\bv X})}\cdot \varphi 
 \\
&= \sum_{i=1}^{Nd}\left [ -\left ( \frac{\partial }{\partial x_i}
\frac{F_i({\bv x})}{\gamma_i}\varphi \right )
+ \frac{T}{\gamma _i}\frac{\partial^2}{\partial x_i^2} \varphi  \right . \\
&\left .  -\frac{F_i({\bv x})}{\gamma _i} \frac{\partial }{\partial X_i} \varphi 
+ \frac{T}{\gamma_i}  \frac{\partial^2}{\partial X_i^2}\varphi  
+\frac{2T}{\gamma _i}  \frac{\partial ^2 }{\partial x_i \partial X_i}
\varphi  \right ].
\label{FPoperator}
\end{split}
\end{equation}
Here, we consider a quantity 
\begin{equation}
q_{\bv h}({\bv x},t) = \int _{{\bf R}^{Nd}} d{\bv X} e^{\sum_{i=1}^{Nd} h_iX_i}p({\bv x},
{\bv X},t),
\label{definitionofq}
\end{equation}
which obeys
\begin{equation}
\frac{\partial q_{\bv h}}{\partial t} 
= \mathcal{L}_{\bv h}^{(\bv x)}\cdot q_{\bv h} ,
\label{timeevolution_q}
\end{equation}
where the operator $\mathcal{L}_{\bv h}^{(\bv x)}$ is calculated as
\begin{equation}
\begin{split}
\mathcal{L}_{\bv h}^{(\bv x)}\cdot \varphi &=\sum_{i=1}^{Nd}
\left [ -\left ( \frac{\partial }{\partial x_i}\frac{F_i({\bv x})}{\gamma_i}
\varphi \right )+ 
\frac{T}{\gamma _i} \frac{\partial^2}{\partial x_i^2} \varphi  \right . \\
&\left . +\frac{F_i({\bv x})}{\gamma _i}h_i \varphi + 
\frac{T}{\gamma _i}h_i^2 \varphi
-\frac{2T}{\gamma _i}h_i 
 \frac{\partial  }{\partial x_i } \varphi  \right ].
\label{modifiedFPoperator}
\end{split}
\end{equation}
Then, from (\ref{XandV}) and (\ref{definitionofq}), we have
\begin{equation}
\langle e^{t \sum_{i=1}^{Nd} h_i   V_i(t)}\rangle 
=\int_{0}^{L}d{\bv x}q_{\bv h} ({\bv x},t).
\label{e^thV}
\end{equation}
Let $\mu_0^{\bv h}$ be 
the largest eigenvalue of $\mathcal{L}_{\bv h}^{(\bv x)}$.
(\ref{timeevolution_q}) and (\ref{e^thV}) lead to
the asymptotic form
\begin{equation}
\langle e^{t \sum_{i=1}^{Nd}h_iV_i(t)}\rangle
\simeq e^{t\mu _0^{\bv h}}
\end{equation}
when $t$ is sufficiently large. 
Thus, from  the definition of $G^{\bv F}(\bv h)$ given 
in (\ref{cumulantgeneratingfunction}), we obtain
\begin{equation}
G^{\bv F}({\bv h}) = \mu ^{\bv h}_0 .
\label{derivation_largesteigenvalue}
\end{equation}
The result that a scaled cumulant generating function is equal to
the largest eigenvalue of an operator is standard \cite{Dembo_Zeitouni}.
The fluctuation theorem was studied with the aid of this result \cite{Lebowitz_Spohn}.
Recently, a relation similar to (\ref{derivation_largesteigenvalue}) was employed for
calculations of the large deviation function of entropy production  \cite{Thomas},
where a kink in the large deviation function has been pointed out.
See Ref. \cite{Svan} for a recent related result.


We denote the adjoint operator of $\mathcal L _{\bv h}^{(\bv x)}$
by $\mathcal L_{\bv h}^{(\bv x)\dagger}$,  which is calculated as
\begin{equation}
\begin{split}
\mathcal{L}_{\bv h}^{(\bv x)\dagger}\cdot \varphi  &=\sum_{i=1}^{Nd} \left [
\frac{F_i({\bv x})}{\gamma_i} 
\frac{\partial }{\partial x_i}
\varphi + \frac{T}{\gamma_i}
 \frac{\partial^2}{\partial x_i^2} \varphi  \right . \\
&  \left . + \frac{F_i({\bv x})}{\gamma_i}h_i \varphi + 
\frac{T}{\gamma _i}h_i^2 \varphi
+\frac{2T}{\gamma_i}h_i 
 \frac{\partial  }{\partial x_i } \varphi  \right ].
\label{modifiedadjointFPoperator1}
\end{split}
\end{equation}
Let $\mu_n^{\bv h}$ and $\eta_n^{\bv h}$ $(n=0,1,2,...)$ be 
all the eigenvalues of $\mathcal L_{\bv h}^{(\bv x)}$
and  $\mathcal L_{\bv h}^{(\bv x)\dagger}$, respectively.
We label the eigenvalues such that
${\rm Re}(\mu_n^{\bv h}) \geq {\rm Re}(\mu_m^{\bv h})$
and ${\rm Re}(\eta_n^{\bv h}) \geq {\rm Re}(\eta_m^{\bv h})$
for $n< m$, where ${\rm Re}(\varphi)$ represents  the real part 
of $\varphi$. We may set in general
\begin{equation}
\mu_n^{\bv h} = (\eta _n^{\bv h})^* 
\end{equation}
with  $n=0,1,2,...$, where  $\varphi^*$ represents 
the complex conjugate of $\varphi$.
As shown in Appendix C, the largest eigenvalues $\mu_0^{\bv h}$ 
and $\eta _0^{\bv h}$ are real, and thus
\begin{equation}
\mu _0^{\bv h} = \eta _0^{\bv h}.
\label{largest_equal}
\end{equation}
Furthermore, the eigenfunctions of the largest eigenvalue
are positive.
From (\ref{derivation_largesteigenvalue}) and (\ref{largest_equal}),
we obtain
\begin{equation}
{\mathcal L} _{\bv h}^{(\bv x)\dagger}\cdot \psi _{\bv h}  
= G^{\bv F}(\bv h) \psi _{\bv h} ,
\label{eigenad}
\end{equation}
where $\psi_{\bv h}({\bv x})$ is the positive eigenfunction 
corresponding to the largest eigenvalue $\eta _0^{\bv h}$.
Here, by noting 
$\log \psi _{\bv h} ({\bv x}) \in {\bf R}$, we define a force
$\bv u_{\bv h}^{\bv F}(\bv x)$ as
\begin{equation}
u_{{\bv h},i}^{\bv F}({\bv x}) 
\equiv 2T\left (\frac{\partial}{\partial x_i}\log \psi _{\bv h} ({\bv x})
 +h_i \right ),
\label{u_hdef}
\end{equation}
which corresponds to the Cole-Hopf transformation \cite{Burgers}.
By using trivial identities
\begin{equation}
\frac{\partial \psi _{\bv h}({\bv x})}{\partial x_i} 
= \psi _{\bv h}({\bv x}) 
\frac{\partial \log \psi _{\bv h} ({\bv x})}{\partial x_i},
\label{cole-hopf-identity1}
\end{equation}
\begin{equation}
\frac{\partial ^2 \psi _{\bv h} ({\bv x})}{\partial  x_i^2} 
= \psi _{\bv h}({\bv x}) 
\left \{ \frac{\partial^2 \log \psi_{\bv h} ({\bv x})}{\partial x_i^2} 
+
\left( 
\frac{\partial \log \psi _{\bv h} ({\bv x})}{\partial x_i} \right )^2 
\right\} 
\label{cole-hopf-identity2},
\end{equation}
we rewrite (\ref{eigenad}) as
\begin{equation}
\begin{split}
G^{\bv F}(\bv h) =\frac{1}{2T}& \sum_{i=1}^{Nd}
\left [\frac{F_i({\bv x})}{\gamma_i}u_{{\bv h},i}^{\bv F}({\bv x})  \right . \\
&\left . + \frac{1}{2\gamma_i} 
u_{{\bv h},i}^{\bv F}({\bv x}) ^2+  \frac{T}{\gamma _i} \frac{\partial}{\partial x_i} 
u_{{\bv h},i}^{\bv F}({\bv x})  \right] .
\label{cumulant}
\end{split}
\end{equation}


Now, we consider a modified system with an arbitrary
external force $\bv w(\bv x)$ added. The stationary probability density
in the modified system, $P_{\rm st}^{\bv F+\bv w}(\bv x)$, 
satisfies (\ref{stationarryrho}) and
(\ref{stationarrycontinue}) with the replacement of $\bv F$ by $\bv F +\bv w$.
By multiplying (\ref{cumulant}) by $P ^{\bv F +  \bv w}_{\rm st}({\bv x})$,
integrating it with respect to $\bv x$, 
and integrating by parts in the third term, we obtain
\begin{equation}
\begin{split}
G^{\bv F}(\bv h) 
=\frac{1}{2T}&\int _{\Omega ^N} d\bv x \sum_{i=1}^{Nd}u_{{\bv h},i}^{\bv F}({\bv x}) 
  \left [ 
P ^{\bv F +  \bv w}_{\rm st}({\bv x})\frac{F_i({\bv x})}{\gamma_i} \right . \\
& \left . + P ^{\bv F +  \bv w}_{\rm st}({\bv x}) \frac{1}{2\gamma_i} 
u_{{\bv h},i}^{\bv F}({\bv x}) -  
\frac{T}{\gamma _i} 
\frac{\partial  P ^{\bv F +  \bv w}_{\rm st}({\bv x})}{\partial x_i} 
\right] .
\label{derivation_cumulant1}
\end{split}
\end{equation}
Then, we substitute (\ref{stationarryrho}) into (\ref{derivation_cumulant1}) 
and we replace
$P ^{\bv F +  \bv w}_{\rm st}({\bv x}) F_i(\bv x)/\gamma_i -(T/\gamma_i) 
\partial P^{\bv F+ \bv w}_{\rm st}/\partial x_i$ 
by $J_{{\rm st},i}^{\bv F+\bv w}(\bv x) - P_{{\rm st}}
^{\bv F}(\bv x)w_i(\bv x)/\gamma_i$.
Furthermore, by substituting (\ref{u_hdef}) into the term $u_{\bv h,i}^{\bv F}(\bv x)J_{{\rm st},i}^{\bv F+\bv w}(\bv x)$,
we rewrite (\ref{derivation_cumulant1}) as
\begin{equation}
\begin{split}
& G^{\bv F}({\bv h}) =  \int _{\Omega ^N} d\bv x\sum_{i=1}^{Nd} \left [ h_i J_{{\rm st},i}^{{{\bv F}+{\bv  w} }}({\bv x})
 - \frac{P_{\rm st}^{{\bv F}+{\bv w}}({\bv x}) }{4T\gamma_i} 
 w_i({\bv x})^2  \right ] \\
& + \int _{\Omega ^N} d\bv x\sum_{i=1}^{Nd} 
\frac{P_{\rm st}^{{\bv F}+{\bv w}}({\bv x}) }{4T\gamma_i}  
\left ( u_{{\bv h},i}^{\bv F}({\bv x}) - w_i({\bv x}) \right )^2 ,
\label{before_G(h)experiment}  
\end{split}
\end{equation}
where we have used (\ref{stationarrycontinue}).
Finally, by recalling (\ref{averaged}) and (\ref{hannkannsuu}), we arrive at
\begin{equation}
G^{\bv F}({\bv h}) = \Phi _{\bv h}^{\bv F}(\bv w)+ \sum_{i=1}^{Nd}
\frac{1}{4T\gamma_i}  
\left \langle \left ( u_{{\bv h},i}^{\bv F} - w_i \right )^2
\right
\rangle^{\bv F +  \bv w} _{\rm st}. 
\label{G(h)experiment}
\end{equation}
It should be noted that the external force $\bv w({\bv x})$
in (\ref{G(h)experiment}) is arbitrary.


Here, the second term of the right-hand side in (\ref{G(h)experiment})
is not negative and becomes zero only when $\bv w = \bv u^{\bv F}_{\bv h}$.
This leads to
\begin{equation}
G^{\bv F}(\bv h) 
= \Phi _{\bv h}^{\bv F}( \bv u_{\bv h}^{\bv F}) 
\geq \Phi _{\bv h}^{\bv F}( \bv w),
\label{derivationofmainclaim1}
\end{equation}
where the equality holds only when $\bv w = \bv u^{\bv F}_{\bv h}$.
This result is the variational expression 
(\ref{cumulantvariational_expression}).
The optimal force $\bv w^{\bv F,\rm opt}_{\bv h} (\bv x)$
defined by (\ref{argment}) is now derived as
\begin{equation}
\bv w^{\bv F,\rm opt}_{\bv h} (\bv x)
= \bv u^{\bv F}_{\bv h}(\bv x) .
\label{derivation_a}
\end{equation}
Next, the derivative of (\ref{G(h)experiment}) with respect to $h_i$ 
yields
\begin{equation}
\begin{split}
&\frac{\partial G^{\bv F}(\bv h)}{\partial h_i} \\
&=  \left \langle V_i \right \rangle_{\rm st}^{\bv F+\bv w}
+ \sum_{i=1}^{Nd}\left [
\frac{ 1}{2T\gamma_i} 
\left \langle \left ( u_{{\bv h},i}^{\bv F} - w_i \right ) 
\label{derivation_b}
\frac{\partial u_{{\bv h},i}^{\bv F}}{\partial h_i}  \right
\rangle^{\bv F +  \bv w} _{\rm st}
\right ]  .
\end{split}
\end{equation}
Since $\bv w$ is arbitrary, we may set
$\bv w = \bv w_{\bv h}^{\bv F,{\rm opt}}$ in the expression. 
This provides the result (\ref{mainclaim_of_the_paper}).
Furthermore, from (\ref{u_hdef}) and (\ref{derivation_a}), we obtain 
(\ref{wopt_psi}) with (\ref{derivation_linear_response1}), because
the eigenfunction $\psi_{\bv h}(\bv x)$ satisfies
$\lim _{\bv h \rightarrow 0}\psi_{\bv h}(\bv x) =\rm const$.


\section{Remarks on the variational function $\Phi_{\bv h}^{\bv F}$}
\label{Section_whatismaximized}

In this section, we present three remarks on 
the variational function $\Phi _{\bv h}^{\bv F}(\bv w)$.
Before entering the main part, we 
first show a slightly different expression of 
$\Phi _{\bv h}^{\bv F}(\bv w)$. 
Since we have (\ref{wopt_psi}),
we may restrict $\bv w (\bv x)$ to the form
\begin{equation}
\bv w({\bv x}) = 
2T\bv h - \frac{\partial }{\partial \bv x} V({\bv x}). 
\label{deltaw_potential}
\end{equation}
Throughout this section, the extra force $\bv w(\bv x)$ 
is always connected to
the potential function $V(\bv x)$ through (\ref{deltaw_potential}).
We then rewrite (\ref{cumulantvariational_expression}) as
\begin{equation}
G^{\bv F}(\bv h) =
\max_{ V}  \Phi _{\bv h}^{\bv F} \left ( \bv w \right )  ,
\label{cumulantvariational_expression2_2}
\end{equation}
where the optimal potential $V^{\bv F,\rm opt}_{\bv h}$ satisfies
\begin{equation}
\bv w^{\bv F,\rm opt}_{\bv h } (\bv x)  
=2T\bv h - \frac{\partial}{\partial \bv x}V^{\bv F,{\rm opt}}_{\bv h} (\bv x). 
\label{wopt_psi_a}
\end{equation}
Furthermore,  by noting
\begin{equation}
\int _{\Omega ^N} d\bv x\sum_{i=1}^{Nd} 2Th_i J_{{\rm st},i} ^{{\bv F}+
 \bv w}({\bv x})
=\int _{\Omega ^N} d\bv x \sum_{i=1}^{Nd} w_i({\bv x}) 
J_{{\rm st},i}^{{\bv F} + \bv w }({\bv x}),
\label{resultofcontinuity}
\end{equation}
and by using (\ref{averaged}), we express
$\Phi _{\bv h}^{\bv F} \left ( \bv w \right )$ as
\begin{equation}
\begin{split}
\Phi _{\bv h}^{\bv F} \left (\bv w \right )=\frac{1}{4T}
   \int _{\Omega ^N} d\bv x\sum_{i=1}^{Nd} &
   { \Bigg [} 2
J_{{\rm st},i}^{{\bv F}+ {\bv w} } ({\bv x}) w_i(\bv x) \\
&  -\frac{P_{\rm st}^{{\bv F}+{\bv w}}({\bv x})}{\gamma_i}
w_i(\bv x)^2
{\Bigg ]}  .
\label{variational_expression1_2}
\end{split}
\end{equation}
The expression of (\ref{cumulantvariational_expression2_2})
with (\ref{variational_expression1_2}) was reported in
Ref. \cite{Nemoto_Sasa}.


This section is organized as follows.
In section \ref{defference_of_entropy_production}, we rewrite the variational
function (\ref{variational_expression1_2}) as the difference between
two entropy production rates. 
In section \ref{relation_to_donsker_varadhan}, 
we point out a similarity between the Donsker-Varadhan formula and our formula.
Finally, in section \ref{relationship_to_the_least},
we show that the cumulant generating function of time-averaged current
in a simple electric circuit
is expressed in terms of the least energy dissipation rate associated with
a variational principle that determines the voltage distribution.


\subsection{Expression of $\Phi _{\bv h}^{\bv F}$ 
in terms of entropy production rate}
\label{defference_of_entropy_production}

In this subsection, we express the variational function (\ref{variational_expression1_2}) in terms of
entropy production rate.
First, the probability current for a given probability density $P(\bv x)$
is written as
\begin{equation}
q_{ i}^{\bv F}({\bv x}|P)  = \frac{F_i(\bv x)}{\gamma_i}P(\bv x)
-\frac{T}{\gamma_i}\frac{\partial P(\bv x)}{\partial x_i},
\label{def_q}
\end{equation}
where the slightly heavy notation $q_{ i}^{{\bv F}}({\bv x}|P)$ is used in order to
distinguish $\bv F$ and $P$ dependence in the current.
Here, we define the functional
$\sigma^{\bv F }(P)$ as
\begin{equation}
\sigma ^{\bv F}(P)  
\equiv  \frac{1}{T}\int _{\Omega ^N} d\bv x\sum_{i=1}^{Nd} \frac{\gamma_i}{P
({\bv x})}\left (q_{ i}^{{\bv F}}({\bv x}|P)  \right )^2.
\label{def_psi}
\end{equation}
 By substituting (\ref{def_q}) into (\ref{def_psi}), we express $\sigma^{\bv F}(P)$
 as
\begin{equation}
\begin{split}
&\sigma^{\bv F} (P)\\
 &= \int _{\Omega ^N} d\bv x\sum_{i=1}^{Nd} q_{ i}^{{\bv F} }({\bv x}|P)  
 \left ( -  \frac{\partial}{\partial x_i}\log P(\bv x)+ \frac{1}{T} F_i(\bv x) \right ).
\end{split}
\end{equation}
The integration by parts of the first term leads to
\begin{equation}
\begin{split}
\sigma ^{\bv F} (P) 
  = &   \int _{\Omega ^N} d\bv x\log P(\bv x) \sum_{i=1}^{Nd}\frac{\partial}{\partial x_i}
 q_i^{\bv F}(\bv x|P)  \\
 &+ \frac{1}{T}
 \int _{\Omega ^N} d\bv x\sum_{i=1}^{Nd} q_{ i}^{{\bv F} }({\bv x}|P)  F_i(\bv x) .
 \label{psi_expression}
\end{split}
\end{equation}
We denote the first and second terms of the right-hand side in (\ref{psi_expression}) by $\Sigma^{\bv F}(P)$ and $\Omega^{\bv F}(P)$, respectively.
Then, $\Sigma ^{\bv F}(P)$ is equal to the increasing rate of the Shannon entropy and
$\Omega^{\bv F}(P)$ corresponds to the energy-dissipation rate divided by $T$.
See Appendix \ref{appendix_entropyproduction} for a more detailed explanation. 
Since $\Omega^{\bv F}(P)$ is identical to the increasing rate of the 
thermodynamic entropy of the heat bath, we identify $\sigma ^{\bv F} (P)$ with
entropy production rate of the total system.

Now, from (\ref{def_q}) and (\ref{def_psi}), we calculate 
\begin{equation}
\begin{split}
&\sigma^{\bv F + \bv w}\left ( P_{\rm st}^{\bv F + \bv w }  \right ) 
- \sigma^{\bv F }\left ( P_{\rm st}^{\bv F + \bv w } \right ) \\
&=\frac{1}{T}\int _{\Omega ^N} d\bv x \sum_{i=1}^{Nd} \Bigg [2q_i^{\bv F+\bv w}(\bv x|P_{\rm st}^{\bv F+\bv w})w_i(\bv x) \\
&\qquad \qquad \qquad \qquad \qquad \qquad - \frac{P_{\rm st}^{\bv F +\bv w}(\bv x)}{\gamma_i} w_i(\bv x)^2 \Bigg ].
\label{sigma_and_q}
\end{split}
\end{equation} 
By noting $q_i^{\bv F+\bv w}(\bv x|P_{\rm st}^{\bv F+\bv w}) = 
J_{{\rm st},i}^{\bv F+\bv w}(\bv x)$ and
comparing (\ref{sigma_and_q}) with (\ref{variational_expression1_2}),
we obtain
\begin{equation}
\Phi _{\bv h}^{\bv F} \left ( \bv w \right )  =\frac{1}{4}
  \left [ \sigma^{\bv F + \bv w}\left ( P_{\rm st}^{\bv F + \bv w }  \right ) 
- \sigma^{\bv F }\left ( P_{\rm st}^{\bv F + \bv w } \right )\right ] .
\label{variational_expression3}
\end{equation}
The first term of the right-hand side in (\ref{variational_expression3}) is
proportional to the entropy production rate of the modified system
in the steady state.
The second term of (\ref{variational_expression3}) is a little bit complicated.
$\sigma^{\bv F}(P_{\rm st}^{\bv F+\bv w})$ is the entropy production rate
just after the extra force $\bv w$ is turned off suddenly
in the steady state of the modified system.
In this manner, the expression (\ref{variational_expression3}) provides us
a thermodynamic interpretation of the variational function $\Phi_{\bv h}^{\bv F}$.


\subsection{Similarity with the Donsker-Varadhan formula}
\label{relation_to_donsker_varadhan}

For a given path $(\bv x(t))_{t=0}^{\tau}$,
we define the occupation density as
\begin{equation}
P_{\tau}^{\rm em}(\bv x) \equiv \frac{1}{\tau}\int _{0}^{\tau} dt \delta(\bv x(t) - \bv x).
\end{equation}
We denote the probability density of $( P_{\tau}^{\rm em}(\bv x))_{\bv x \in \Omega}$
by $\rm Prob_{\tau}^{\bv F}(P)$. We write it formally as 
\begin{equation}
{\rm Prob}_{\tau} ^{\bv F}(P) = \left \langle \prod _{\bv x \in \Omega} \delta\left ( P_{\tau}^{\rm em}(\bv x) -  P(\bv x) \right ) \right \rangle.
\label{path_probability4}
\end{equation}
When $\tau$ is sufficiently large, the asymptotic form of
${\rm Prob}_{\tau}^{\bv F}(P)$ is given by
\begin{equation}
{\rm Prob}_{\tau} ^{\bv F}(P) \simeq e^{-\tau I^{\bv F}(P)},
\end{equation}
which is the large deviation property of the occupation density
with the large deviation function $I^{\bv F}(P)$.
Let
$\mathcal L_{\rm FP}^{(\bv x)\dagger}$ be the adjoint Fokker-Planck operator
defined by
\begin{equation}
\mathcal L_{\rm FP}^{(\bv x)\dagger}\cdot \varphi = 
\sum_{i=1}^{Nd}\left [\frac{F_i(\bv x)}{\gamma_i}  \frac{\partial}{\partial x_i}
\varphi +\frac{T}{\gamma_i} 
 \frac{\partial^2}{\partial x_i^2}\varphi \right ].
\label{adjoin_Fokker_Planck_DV}
\end{equation}
Then, the Donsker-Varadhan formula
in the Langevin system is written as
\begin{equation}
I^{\bv F}(P) = -\min _{\phi >0} \int _{\Omega ^N} d\bv x P(\bv x) \frac{\mathcal L_{\rm FP}^{(\bv x)\dagger}\cdot \phi }{\phi(\bv x)}. \label{DV_variational}
\end{equation}
Such a formula was rigorously derived for general Markov stochastic 
processes \cite{Donsker_Varadhan}.
Its universal form suggests that it might be 
connected to a framework of non-equilibrium statistical mechanics. Indeed, in 
Ref. \cite{Maes_Minimumentropy},
it has been  pointed out that the minimum entropy 
production principle is related to the Donsker-Varadhan formula (\ref{DV_variational}). 
This relation was obtained by noting the detailed balance condition 
in equilibrium systems. Similarly, in equilibrium dynamics, 
the variational principle determining the activity was recognized as a form similar
to the Donsker-Varadhan formula \cite{garrahan}. More recently, even without the detailed balance condition, the variational function in the Donsker-Varadhan
formula was conjectured to be related to the activity \cite{Maes_Linear_Response}.

Based on these achievements, we rewrite (\ref{DV_variational})
so as to see a similarity with our
variational function (\ref{variational_expression3}).
First, since $\phi$ in (\ref{DV_variational}) is positive,
we define the potential $V(\bv x)$ by
\begin{equation}
V(\bv x) \equiv - 2T \log \phi(\bv x).
\end{equation}
Then, using
(\ref{cole-hopf-identity1}) and (\ref{cole-hopf-identity2}), we rewrite (\ref{DV_variational})
as
\begin{equation}
\begin{split}
& I^{\bv F}(P)  = - \frac{1}{2T}\min _{\phi (\bv x)>0} \int _{\Omega ^N} d\bv x P(\bv x) \sum_{i=1}^{Nd} \\ 
&\left [ -\frac{F_i(\bv x)}{\gamma_i}  \frac{\partial V(\bv x)}{\partial x_i}
  -\frac{T}{\gamma_i} 
 \frac{\partial^2 V(\bv x)}{\partial x_i^2}  + \frac{1}{2\gamma_i}
\left (-\frac{\partial V(\bv x)}{\partial x_i} \right )^2 \right ]. 
\label{relation_DV4}
\end{split}
\end{equation}
The integration by parts in the second term leads to
\begin{equation}
\begin{split}
& I^{\bv F}(P)  = - \frac{1}{2T} \min _{\phi (\bv x)>0} \int _{\Omega ^N} d\bv x  \sum_{i=1}^{Nd}
\left (-\frac{\partial V(\bv x)}{\partial x_i} \right )  \\
& \left [\frac{F_i(\bv x)}{\gamma_i}P(\bv x) - \frac{T}{\gamma_i} 
\frac{\partial P(\bv x)}{\partial x_i}  + \frac{1}{2\gamma_i}
\left (-\frac{\partial V(\bv x)}{\partial x_i} \right ) P(\bv x)\right ]. 
\end{split}
\end{equation}
Thus, by using (\ref{def_q}) with the replacement of $\bv F$ by 
$\bv F - \partial V/\partial \bv x$, we obtain
\begin{equation}
\begin{split}
& I^{\bv F}(P)  = -\frac{1}{4T}\min _{V(\bv x)} \int _{\Omega ^N} d\bv x  \sum_{i=1}^{Nd} \\
& \left [ 2\left (-\frac{\partial V(\bv x)}{\partial x_i}  \right ) q_i^{\bv F -\partial V/\partial \bv x}(\bv x|P)  - \frac{1}{\gamma_i}
\left (-\frac{\partial V(\bv x)}{\partial x_i} \right ) ^2P(\bv x) \right ].
\label{relation_DV2} 
\end{split}
\end{equation}
Furthermore, by comparing (\ref{relation_DV2}) with (\ref{sigma_and_q}),
we find
\begin{equation}
I^{\bv F}(P)  = -\frac{1}{4}\min _{V} \int _{\Omega ^N} d\bv x  
\left [ \sigma^{\bv F -\partial V/\partial \bv x}\left ( P \right ) 
- \sigma^{\bv F }\left ( P \right ) \right ].
\label{relation_DV3}
\end{equation}
These expressions, (\ref{relation_DV2}) and (\ref{relation_DV3}), 
were derived in Ref. \cite{Mars_similarity, large-deviation-technique} by using another method. 
For the large deviation function of occupation density in jump processes, a similar expression in terms of escape rates
instead of the entropy production rates has been reported in Ref.
\cite{Maes_Linear_Response}.

It is seen that there is a strong similarity between (\ref{relation_DV3}) and 
(\ref{cumulantvariational_expression2_2})
with (\ref{variational_expression3}).
This suggests that two formulas might be described in
a unified manner. 
We will report a unified treatment of the
Donsker-Varadhan formula and ours in another paper.


\subsection{Relationship to the principle of the least dissipation of energy}
\label{relationship_to_the_least}

\begin{figure}[tbh]
\includegraphics[width=8cm]{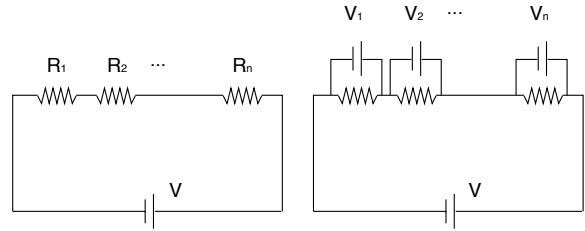}
\caption{Left: $n$ resistances ($R_1, R_2, ..., R_n$)
are connected in series. We impose an electric potential $V$ on the circuit. Right: By using $n$ batteries, we impose an electric potential $V_i$ on each resistance $R_i$.}
\label{fig1}
\end{figure}

We consider an  electric circuit in which $n$ resistances ($R_1, R_2,...,R_n$)
are connected in series.
We impose an electric potential $V$ on the circuit 
by using a battery as shown in the left side of Fig. \ref{fig1}.

Let $V_{*i}$ be the difference of the electric potential at the resistance $R_i$
in the electric circuit. This $V_{*i}$ is determined by Ohm's law
\begin{equation}
V_{*i} = I_* R_i,
\label{Ohm's_law}
\end{equation}
where $I_*$ is the electric current in the circuit. From the constraint condition $\sum_{i=1}^{n}V_{*i}=V$ and Ohm's law (\ref{Ohm's_law}), we obtain
\begin{equation}
I_*=\frac{V}{\sum_{i=1}^{n}R_i},
\end{equation} 
\begin{equation}
V_{*i}=V\frac{R_i}{\sum_{j=1}^nR_j}.
\label{voltage}
\end{equation}

The principle of the least dissipation of energy claims that
$V_{*i}$ is characterized by
the least dissipation of energy \cite{ono_review}.
In order to see this statement,
we consider an electric circuit 
displayed in the right side of Fig. \ref{fig1}.
By using $n$ batteries, we impose an electric potential $V_i$ on each resistance $R_i$ such that $\sum_{i=1}^{n}V_i=V$.
Under the assumption that the energy dissipation is determined by the Joule heating,
the energy dissipation rate $K$ is 
calculated as
\begin{equation}
K(V_1,V_2,...,V_n) = \sum _{i=1}^n \frac{V_i^2}{R_i}.
\end{equation}
Then,
one can directly confirm
that $V_{*i}$ is
determined as
\begin{equation}
(V_{*i})_{i=1}^n =   \underset{\underset{(\sum_iV_i=V)}{(V_i)_{i=1}^{n}}}{\rm argmin }
K(V_1,V_2,...,V_n) ,
\end{equation}
where $(\sum_{i=1}^n V_i = V)$ means the constraint condition for
$(V_i)_{i=1}^{n}$.
This is an example of the principle of the least dissipation of energy.
We define the least energy dissipation rate $K_*(V)$ as
\begin{equation}
K_*(V) =   \underset{\underset{(\sum_iV_i=V)}{(V_i)_{i=1}^{n}}}{\rm min }
 K(V_1,V_2,...,V_n) .
 \label{leastenergydissipationrate}
\end{equation}

Next, we consider fluctuations of the time-averaged electric current $I$ during
a time interval $[0,\tau]$.
In a manner similar to (\ref{cumulantgeneratingfunction}), we define the scaled
cumulant generating function of the time-averaged current as
\begin{equation}
G(h) \equiv \lim_{\tau \rightarrow \infty} \frac{1}{\tau} \log \left \langle e^{\tau I h}\right 
\rangle.
\end{equation}
For simplicity, we assume that the current fluctuations are described by Johnson noise.
See the first chapter of Ref. \cite{Gardiner} for a brief review of Johnson noise. The 
Johnson noise
satisfies the fluctuation dissipation relation
\begin{equation}
\lim _{\tau \rightarrow \infty}\tau \left (  \langle I^2 \rangle - \langle I \rangle ^2\right ) = \frac{2T}{\sum_{i}R_i}.
\end{equation}
We also assume that the fluctuations are Gaussian.
Under these assumptions, we write $G(h)$ as
\begin{equation}
G(h) = \frac{V}{\sum_{i=1}^n R_i} h + \frac{T}{\sum_{i=1}^nR_i} h^2.
\label{electric_cumulantgenerating}
\end{equation}

Now, we connect the least energy dissipation rate $K_*(V)$ and the cumulant generating function $G(h)$.
First, from (\ref{voltage}) and (\ref{leastenergydissipationrate}), we find
\begin{equation}
K_*(V)
= \frac{V^2}{\sum_{j}R_j}. \label{leastdissipation}
\end{equation}
Then, by replacing $V$ by $V+2Th$ in (\ref{leastdissipation}), we obtain
\begin{equation}
K_*(V+2Th)
= \frac{V^2}{\sum_{j}R_j}+4T G(h) ,  \label{leastdissipation2}
\end{equation}
where we have used (\ref{electric_cumulantgenerating}).
The subtraction of (\ref{leastdissipation}) from
(\ref{leastdissipation2}) leads to
\begin{equation}
G(h) = \frac{1}{4T}\left [ K_*(V+2Th) - K_*(V) \right ].
\label{leastdissipation_difference}
\end{equation} 
This result that $G(h)$ is written by the difference of the energy dissipation
rates for two different states reminds us of (\ref{cumu_vs_gibbs})
in equilibrium statistical mechanics.
Our formula (\ref{cumulantvariational_expression2_2})
with (\ref{variational_expression3}) is a generalization of 
(\ref{leastdissipation_difference}).


\section{Applications}
\label{Section_application}

In this section, we present three applications of our
result by restricting our investigation 
to a Brownian particle on a ring. We write an explicit
setting. Let $x(t)$ be a position of the particle along 
the ring of size $L$. We assume that 
the motion of the particle is described by
\begin{equation}
\dot x = \frac{1}{\gamma}F(x) + \sqrt{\frac{2T}{\gamma}}\xi 
\label{1langevin}
\end{equation}
with the periodic boundary condition, 
where $F(x)$ is a deterministic force given by
\begin{equation}
F(x)=f-\frac{\partial U(x)}{\partial x}. \label{1langevin_force}
\end{equation}
Such a force can be realized by using a rotating periodic 
potential \cite{Bechinger, Ciliberto}. Note that (\ref{1langevin}) with 
(\ref{1langevin_force}) corresponds to the case that $N=d=1$
in the model studied in Section \ref{section_model}.
The stationary distribution $P_{\rm st}^F(x)$ and the stationary
current $J_{\rm st}^F$ are determined by
\begin{equation}
J_{\rm st}^F = \frac{1}{\gamma}F(x)P_{\rm st}^F(x)
-\frac{T}{\gamma} \frac{\partial}{\partial x} P_{\rm st}^F(x) 
\label{1d-statonarrycurrent}
\end{equation}
with the periodic boundary condition $P_{\rm st}^F(0)=P_{\rm st}^F(L)$
and the normalization condition $\int _{0}^L dxP_{\rm st}^F(x)=1$.

For this system, in section \ref{Section_diffusionconstant},
we derive the diffusion constant 
$D=\lim _{\tau \rightarrow \infty}(1/2)\tau\left \langle 
V(\tau) ^2\right \rangle_c$ and the mobility 
$\mu = \left ( \partial /\partial f\right )
\lim _{\tau \rightarrow \infty} \left \langle 
V(\tau) \right \rangle_c$ by utilizing our formula.
Then, in section \ref{appendix_additivity},
we show that our variational expression (\ref{cumulantvariational_expression})
with (\ref{hannkannsuu}) leads to a formula derived from
an additivity principle \cite{BD, BD2, BD3, Derrida_review}. Finally,
in section \ref{Section_measurement}, we present an experimental
method for determining the cumulant generating function 
without the information of $F(x)$.


\subsection{Diffusion constant and mobility}
\label{Section_diffusionconstant}

In the system under study, 
the stationary distribution $P_{\rm st}^F(x)$, the diffusion
constant $D$ and the mobility $\mu$ are known to be expressed in terms
of the following functions \cite{Reimann, Reimann2, Hayashi-Sasa}:
\begin{equation}
I_{\pm}(x) \equiv \int _{0}^{L}dy e^{\pm  \int _{x}^{x  \mp y}dz F(z)/T }.
\label{Iplusminus}
\end{equation} 
Concretely, the expressions are
\begin{equation}
P_{\rm st}^F(x) = \alpha I_{-}(x),
\label{remark_reimann_4}
\end{equation}
\begin{equation}
D=\frac{T}{\gamma}L^2\alpha^3 \int_{0}^L dx I_{+}(x) I_{-}(x)^2 ,
\label{remark_reimann_9}
\end{equation}
and
\begin{equation}
\mu =\frac{L}{\gamma}\alpha^2 \int _{0}^L dx I_+(x)I_-(x),
\label{remark_derivative_3_a}
\end{equation}
where $\alpha$ is the normalization constant given by
\begin{equation}
\alpha \equiv \left ( \int _{0}^Ldx I_{\pm}(x)\right )^{-1} . 
\label{Ipm_alpha}
\end{equation}
Here, one may wonder about the origin of such compact expressions.
Below, within
our formulation, we uncover a simple structure behind the expressions.

First, we consider the diffusion constant. We start with the formula (\ref{G(h)experiment}) with (\ref{hannkannsuu}).
In the case that $N=d=1$, it becomes
\begin{equation}
\begin{split}
&G^{F}(h) \\
& = h\left \langle V\right \rangle_{\rm st}^{F+w} - \frac{1}{4T\gamma} \left [
\left \langle
w^2 \right \rangle_{\rm st}^{F+w} - \left \langle \left ( u_h^F - w \right )^2\right 
\rangle _{\rm st}^{F+w}\right ], \label{remarks_reimann_1}
\end{split}
\end{equation}
where it should be noted that $w(x)$ is arbitrary. By setting
$w=0$ in (\ref{remarks_reimann_1}), we obtain
\begin{equation}
G^{F}(h) = h\left \langle V\right \rangle_{\rm st}^{F} + \frac{1}{4T\gamma}  \left \langle  (u_h^F) ^2\right \rangle _{\rm st}^{F}.
\label{remarks_reimann_2}
\end{equation}
Since $u_{h}^F(x) |_{h=0}= 0$ (See (\ref{derivation_linear_response1}) and (\ref{u_hdef})), $u_h^F(x)$ is expanded as
\begin{equation}
u_h^F(x) = \sum_{n=1}^{\infty} u_n(x) h^n.
\label{remarks_reimann_3}
\end{equation}
The substitution of (\ref{remarks_reimann_3}) into (\ref{remarks_reimann_2})
yields
\begin{equation}
G^{F}(h) = h\left \langle V\right \rangle_{\rm st}^{F} + \frac{1}{4T\gamma}  \left \langle  (u_1) ^2\right \rangle _{\rm st}^{F} h^2 + \mathcal O(h^3),
\label{remark_reimann_8}
\end{equation}
which leads to
\begin{equation}
D= \frac{1}{4T\gamma}  \left \langle  (u_1) ^2\right \rangle _{\rm st}^{F} .
\label{remark_reimann_6}
\end{equation}

Next, we consider the mobility. Let us fix a function $w^\prime(x)$. By setting
$w=-F+w^\prime$ in (\ref{remarks_reimann_1}), we rewrite (\ref{remarks_reimann_1}) as
\begin{equation}
G^F(h) = h\left \langle V \right \rangle _{\rm st}^{w^\prime}
-\frac{1}{4T\gamma}\left \langle (w^\prime - F)^2 
-(u_{h}^F - w^\prime +F)^2 \right \rangle _{\rm st}^{w^\prime}.
\label{remarks_derivatives_1}
\end{equation}
We then differentiate (\ref{remarks_derivatives_1}) with respect to force $f$.
The result is
\begin{equation}
\begin{split}
& \frac{\partial G^F(h)}{\partial f}  \\
&= \frac{1}{2T\gamma}\left \langle (w^\prime - F)
+(u_{h}^F - w^\prime +F)\left (\frac{\partial u_h^F}{\partial f}+1 \right ) \right \rangle _{\rm st}^{w^\prime}.
\end{split}
\end{equation}
Since $w^{\prime}(x)$ is arbitrary in this expression, we may set
$w^\prime = u_h^F+F$. This leads to
\begin{equation}
\frac{\partial G^F(h)}{\partial f} = \frac{1}{2T\gamma} \left \langle u_h^F \right \rangle_{\rm st}^{F+u^{F}_h}.
\label{remarks_derivatives_2}
\end{equation}
By substituting (\ref{remarks_reimann_3}), (\ref{remark_reimann_8})
into (\ref{remarks_derivatives_2}), we obtain
\begin{equation}
\mu=\frac{1}{2T\gamma} \left \langle  u_1\right  \rangle _{\rm st}^{F}.
\label{remark_derivative_3}
\end{equation}

As shown in (\ref{remark_reimann_6}) and (\ref{remark_derivative_3}),
$D$ and $\mu$ are expressed by the second and the first moments of $u_1(x)$, respectively. Here, we determine $u_1(x)$.
We substitute (\ref{remarks_reimann_3}) into (\ref{cumulant}) and
extract the terms proportional to $h$.
We then obtain
\begin{equation}
2T\left \langle V\right \rangle _{\rm st}^F=\frac{T}{\gamma} \frac{\partial}{\partial x} u_1(x) +\frac{1}{\gamma}F(x)u_1(x).
\label{determination_of_u1}
\end{equation}
Note the periodic boundary condition $u_1(0)=u_1(L)$ and the normalization
condition
\begin{equation}
\int _{0}^{L}dx u_1(x) = 2TL,
\end{equation}
which comes from (\ref{u_hdef}). By comparing (\ref{determination_of_u1})
with (\ref{1d-statonarrycurrent}), we find that
$u_1(x)$ can be solved in the manner similar to $P_{\rm st}^F(x)$.
The result is
\begin{equation}
u_1(x)=2TL\alpha I_{+}(x).
\label{remark_reimann_5}
\end{equation}
By substituting (\ref{remark_reimann_5}) and (\ref{remark_reimann_4})
into (\ref{remark_reimann_6}) and
(\ref{remark_derivative_3}), we obtain (\ref{remark_reimann_9}) and
(\ref{remark_derivative_3_a}).
Furthermore, we can systematically evaluate higher-order cumulant coefficients
by formulating an iterative relation. 
See Appendix \ref{appendix_iteratibe_expression}.

At the end of this subsection, we study the violation of the fluctuation-dissipation
relation.
First, from (\ref{stationary3}), we have 
\begin{equation}
\left \langle V\right \rangle _{\rm st}^{F+u_h^F}-\frac{1}{\gamma} \left \langle F\right \rangle_{\rm st}^{F+u^F_h} = \left \langle u^F_h \right \rangle_{\rm st}^{F+u^F_h}.
\label{disccusion_of_FDT}
\end{equation}
We substitute (\ref{disccusion_of_FDT}) into (\ref{remarks_derivatives_2})
and combine it with our main result (\ref{mainclaim_of_the_paper}). We
then obtain
\begin{equation}
\frac{\partial G^F(h)}{\partial f} = \frac{1}{2T}\frac{\partial G^F(h)}
{\partial h} -  \frac{1}{2T\gamma}\left \langle F \right \rangle_{\rm st}^{F+u^{F}_h},\label{disccusion_of_FDT2}
\end{equation}
which leads to
\begin{equation}
\frac{D}{T} - \frac{\partial \left \langle V\right \rangle _{\rm st}^{F}}{\partial f} = 
\frac{1}{2T\gamma}\lim_{h\rightarrow 0} \frac{\partial }{\partial h} \left \langle F \right \rangle_{\rm st}^{F+u^{F}_h}.
\label{discussion_of_FDT3}
\end{equation}
Furthermore, by considering an expansion form
\begin{equation}
P_{\rm st}^{ F + u_{ h}^{ F}}( x) = P_{\rm st}^{ F}( x)
+ hQ_1(x) + \mathcal O(h^2),
\label{Pexpansion}
\end{equation}
we rewrite the right-hand side of (\ref{discussion_of_FDT3}) as
\begin{equation}
\frac{1}{2T\gamma} \int_{0}^L dx F(x)Q_1(x). 
\label{Degree_of_violation}
\end{equation}
The expression (\ref{Degree_of_violation}) represents a degree of the violation
of the fluctuation-dissipation relation. Recently, several attempts have been presented so as to characterize the violation of fluctuation-dissipation relation
\cite{Harada-Sasa, Maes_violation_of_FDT, Speck_Seifert_EPL, Gawedzki, Prost}.
It might be interesting to find a connection of (\ref{Degree_of_violation})
with them.

When the driving force $f$ is absent,
$Q_1(x)$ becomes zero so that the fluctuation-dissipation relation holds. We show this fact explicitly.
We substitute (\ref{Pexpansion}) into
(\ref{1d-statonarrycurrent}) with the replacement
of $F(x)$ with $F(x)+u_h^F(x)$, and extract the terms proportional to $h$.
We then obtain the following differential equation for $Q_1(x)$:
\begin{equation}
\frac{\partial}{\partial x}\left [
\frac{1}{\gamma} F(x) Q_1(x)
+\frac{1}{\gamma}u_1(x) P_{\rm st}^{F}(x) 
-\frac{T}{\gamma} \frac{\partial}{\partial x}Q_1(x)
\right ]
=0.
\label{Q_1determine}
\end{equation}
with the periodic boundary condition $Q_1(0)=Q_1(L)$ and the normalization condition
\begin{equation}
\int _{0}^Ldx Q_1(x)=0,
\label{Q_1_normalization}
\end{equation}
which comes from (\ref{Pexpansion}).
Here, from (\ref{remark_reimann_4}) and (\ref{remark_reimann_5}),
we find that $u_1(x)P_{\rm st}^F(x)$ is constant in $x$ when
$F(x)=-\partial U(x)/\partial x$. 
In this case, (\ref{Q_1determine}) provides us $Q_1(x) = C_1e^{- U(x)/T}$, where $C_1$
is a constant. Then, (\ref{Q_1_normalization}) leads to $C_1=0$.
Thus, recalling (\ref{Degree_of_violation}), we conclude that the right-hand 
side of (\ref{discussion_of_FDT3}) is equal to zero when $f$ is absent.
Note that the discussion above applies to
many-body systems and higher-dimensional cases.


\subsection{Relation to an additivity principle}
\label{appendix_additivity}
We start with the variational expression (\ref{cumulantvariational_expression}) 
with (\ref{hannkannsuu}).
Since the probability current
 is independent of $x$ in the one-dimensional case,
 we can rewrite (\ref{cumulantvariational_expression}) as
\begin{equation}
G^F(h)=\max_{w} \left [ h J_{\rm st}^{{ F+ w }}L
 - \int_{0}^{L} d x  \frac{P_{\rm st}^{{ F}+ w}({ x}) }{4T\gamma} 
w({ x})^2  \right ] ,
\label{relation_additivity1}
\end{equation}
where $J_{\rm st}^{F+w}$ and $P_{\rm st}^{F+w}(x)$ are the stationary probability
current and density in the modified system. 
Note that $J_{\rm st}^{F+w}$ and $P_{\rm st}^{F+w}(x)$
can be determined uniquely from (\ref{1d-statonarrycurrent}) for a given
extra force $w(x)$. Conversely, when a constant current $J$ and a probability density function $P(x)$ 
are given, we can define the extra force $w(x)$ by
\begin{equation}
w(x) = \frac{\gamma}{P(x)} \left ( J+\frac{T}{\gamma}
 \frac{\partial P(x)}{\partial x} - \frac{P(x)}{\gamma}F(x) \right ).
\end{equation}
Thus, there is a one-to-one correspondence between
$w(x)$ and $\left (J_{\rm st}^{F+w}, P_{\rm st}^{F+w}(x) \right)$.
From this fact, we can rewrite (\ref{relation_additivity1})
as
\begin{equation}
\begin{split}
 &G^F(h) = \max_{ P,J} \Bigg [ h
 JL \\
 &-\frac{1}
 {4T} \int _{0}^L dx \frac{\gamma}{P(x)}\left ( J+\frac{T}{\gamma}
 \frac{\partial P(x)}{\partial x} - \frac{P(x)}{\gamma}F(x) 
\right )^2 \Bigg ].
\end{split}
\end{equation}
By taking the maximum of $P(x)$ first, we obtain
\begin{equation}
\begin{split}
& G^F(h)=\max_{ J}   \Bigg [ hJL \\
&-\frac{1}
 {4T}\min_{P}\left [ \int _{0}^Ldx \frac{\gamma}{P(x)}\left ( J+\frac{T}{\gamma}
 \frac{\partial P(x)}{\partial x} - \frac{P(x)}{\gamma}F(x) 
\right )^2 \right ] \Bigg ]. 
\end{split}
\end{equation}
Therefore, we find that the Legendre transform of
$G^F(h)$ with respect to $hL$, which is the large deviation function $I(J)$, becomes
\begin{equation}
\begin{split}
& I(J) \\
& =\frac{1}
 {4T}\min_{P}\left [ \int _{0}^Ldx \frac{\gamma}{P(x)}\left ( J+\frac{T}{\gamma}
 \frac{\partial P(x)}{\partial x} - \frac{P(x)}{\gamma}F(x)  \label{additivity}
\right )^2 \right ]. 
\end{split}
\end{equation}
This expression is the same as the variational expression derived from
the additivity principle \cite{BD, BD2, BD3, Derrida_review}.

When the system is described by fluctuating hydrodynamics,
the variational expression
is obtained under the assumption that the most dominant contribution to a path integration
is given as a stationary configuration
\cite{phase-transition1, BDGJL, BDGJL2, phase-transition2, phase-transition4}. 
Here, since the Langevin equation (\ref{langevin}) can be mapped to a
stochastic partial differential equation \cite{Dean},
which is equivalent to fluctuating hydrodynamics, 
we can derive the variational expression in our system if the assumption
is satisfied.
Our result (\ref{additivity}) indicates that the assumption is valid.

In the derivation of (\ref{additivity}),
the spatial homogeneity of the stationary current
plays an important role. However, since this property is specific to
one-dimensional systems, 
it is not obvious  whether the same variational principle does hold in
higher-dimensional cases \cite{2dimension}. 
Furthermore, by noticing that the probability current and
the probability density considered here are equivalent
to the particle current and the particle density in non-interacting
particle systems, one might find
that it is a highly non-trivial problem to derive (\ref{additivity})
for interacting particle systems.


\subsection{Experimental determination of cumulant 
generating function}\label{Section_measurement}

We consider an experimental method for determining
$G^F(h)$ without the information of $F(x)$.
Since the direct evaluation of cumulants becomes harder and harder
as the order of the cumulants increases, it may be useful if we could
present an efficient method for evaluating $G^F(h)$
on the basis of our formula under the assumption that we measure
only the trajectories $(x(t))_{t=0}^{\tau}$ and values of $T$ and $\gamma$.

The basic idea is to utilize (\ref{G(h)experiment}), which was
a cornerstone in the derivation of the main results (\ref{cumulantvariational_expression}) and (\ref{mainclaim_of_the_paper}).
First, we add some external forces $w(x)$ to the system and measure trajectories
of the particle.
We then notice that the left-hand side of (\ref{G(h)experiment})
is independent of external forces $w(x)$.
Since we can estimate the right-hand side of (\ref{G(h)experiment})
by using the measured data,
we obtain equations which yield $w^{F,\rm opt}_h(x)$.
Essentially the same idea was reported in Ref. \cite{Nemoto_Sasa}.
Below, we present a slightly different one from that in Ref. \cite{Nemoto_Sasa}.

Concretely, in terms of a complete set of orthogonal functions $(\phi _{\alpha})_{\alpha=0}^{\infty}$,
we expand
$w^{F,\rm opt}_{ h} ( x)$ as
\begin{equation}
 w^{F,\rm opt}_{ h} ( x)\simeq \sum_{\alpha=0}^{m}u^h_\alpha\phi_\alpha(x), \label{expansion1}
\end{equation}
where $m$ is a truncation number of the expansion.
As $m$ is increased, the accuracy of the expansion (\ref{expansion1})
increases. For later convenience, we choose $\phi_0(x)=1$.
Since
$\int_{0}^L dx w_h^{F,\rm opt}(x)=2ThL$ 
(See (\ref{wopt_psi})), the coefficient $u_0^h$ is
derived as
\begin{equation}
u_0^h=2Th . \label{conditionua_0}
\end{equation}
We shall determine the other expansion coefficients $(u_{\alpha}^h)_{\alpha=1}
^m$.

First, by differentiating (\ref{G(h)experiment}) with respect to $h$, we obtain
\begin{equation}
\frac{\partial G^F(h)}{\partial h} =\left \langle V\right \rangle_{\rm st}^{F+ w} +
 \frac{1}{2T\gamma} \left \langle \left (
w^{F,\rm opt}_h -  w
\right ) \frac{\partial  w^{\rm opt}_h}{\partial h}  \right \rangle^{ F+ w}_{\rm st} 
\label{1G(h)}.
\end{equation}
We then expand $w(x)$ as
\begin{equation}
w(x) = \sum_{\alpha=0}^{m}w_\alpha \phi_\alpha(x). \label{expansion2}
\end{equation}
By substituting (\ref{expansion1}) and (\ref{expansion2}) into (\ref{1G(h)}), we
rewrite (\ref{1G(h)}) as
\begin{equation}
\begin{split}
&\frac{\partial G^F(h)}{\partial h} \\
&=\left \langle V\right \rangle^{F+w} _{\rm st}
+\frac{1}{2T\gamma} \sum_{\alpha=0}^m \sum_{\eta=0}^m 
\left ( u_\alpha ^h -
w_\alpha \right )
\left \langle \phi_\alpha  \phi_\eta   \right \rangle^{ F+w}_{\rm st}
\frac{\partial u_\eta^h}{\partial h}.
\label{1experiment1}
\end{split}
\end{equation}
Here, by defining
\begin{equation}
A_{\eta } ^{w} \equiv \frac{1}{2T\gamma}\sum_{\alpha =0}^m(u_\alpha^h - w_\alpha)
\langle \phi_\alpha \phi_\eta \rangle ^{F+w}_{\rm st},
\end{equation}
we rewrite (\ref{1experiment1}) as
\begin{equation}
\frac{\partial G^F (h)}{\partial h} =\left \langle V\right \rangle^{F+ w}_{\rm st} 
+ 2TA_0^{w}+
\sum_{\eta=1}^m A_\eta ^{w}\frac{\partial u_{\eta}^{h}}{\partial h}
\label{Gdiffexperiment},
\end{equation}
where we have used (\ref{conditionua_0}). 
We prepare $m+1$ external forces $w^{(0)}(x),w^{(1)}(x),...,w^{(m)}(x)$.
Recalling that (\ref{Gdiffexperiment}) is valid for any external force $w(x)$,
we equate the right-hand side of (\ref{Gdiffexperiment}) for the case
$w^{(0)}(x)$ to the right-hand side of (\ref{Gdiffexperiment})
for the other cases
$w^{(\alpha)}(x)$ $(\alpha=1,...,m)$.
The obtained equations are
\begin{equation}
\begin{split}
& \left \langle V\right \rangle^{F+w^{(0)}}_{\rm st}+2TA_0^{w^{(0)}}
 -\left \langle V\right \rangle^{F+w^{(\alpha)}}_{\rm st}-2TA_0^{w^{(\alpha)}} \\
 & =  \sum_{\eta =1}^m \left ( A_\eta ^{w^{(\alpha)}}-
A_\eta^{w^{(0)}} \right ) \frac{\partial u_\eta ^h}{\partial h},
\label{Gdiffexperiment1}
\end{split}
\end{equation}
where $\alpha=1,...,m$. 
We further define an $m\times m$ matrix as 
\begin{equation}
\bar A _{\alpha \eta}\equiv 
A_\eta ^{w^{(\alpha)}}-
A_\eta^{w^{(0)}}.
\end{equation}
With the assumption that $\bar A$ is invertible, we derive
\begin{equation}
\begin{split}
 \frac{\partial u_\eta ^h}{\partial h}
=\sum_{\alpha =1}^{m}& (\bar A^{-1})_{\eta \alpha}    \left [ 
\left \langle V\right \rangle^{F+w^{(0)}}_{\rm st}+2TA_0^{w^{(0)}} \right . \\
& \left .  -\left \langle V\right \rangle^{F+w^{(\alpha)}}_{\rm st}-2TA_0^{w^{(\alpha)}}
\right ].
\label{rennritu}
\end{split}
\end{equation}

This expression implies that
$\partial u_{\eta}^h/\partial h$ is expressed in terms of
$(u_{\eta}^h)_{\eta=1}^{m}$, $\left \langle V \right \rangle_{\rm st}^{F+w^{(\alpha)}}$
and $\left \langle \phi_\eta\phi_\delta \right \rangle _{\rm st}^{F+w^{(\alpha)}}$.
Now, from trajectories of the particle in the modified systems with the external
forces $w^{(\alpha)}(x)$ added,
we approximately obtain
$\left \langle V \right \rangle_{\rm st}^{F+w^{(\alpha)}}$
and $\left \langle \phi_\eta\phi_\delta \right \rangle _{\rm st}^{F+w^{(\alpha)}}$
in the form
\begin{equation}
\left \langle V\right \rangle^{F+w^{(\alpha)}}_{\rm st} \simeq \frac{1}{\tau} \int _{0}^{\tau}dt
 \dot x(t),
 \label{apploxi_V}
\end{equation}
\begin{equation}
\left \langle \phi_\eta\phi_\delta \right \rangle _{\rm st}^{F+w^{\alpha}} 
\simeq \frac{1}{\tau}\int _{0}^{\tau}dt \phi_\eta  \left (x(t) \right )\phi_
\delta  \left ( x(t) \right ),
\label{apploxi_phi} 
\end{equation}
with $0\leq \eta$, $\delta \leq m$. Note that the approximation
becomes exact in the limit $\tau \rightarrow \infty$.
Therefore, we can express
$\partial u_{\eta}^h/\partial h$ by 
$(u_{\eta}^h)_{\eta=1}^{m}$ from experimental data.
By integrating $\partial u_\eta ^h / \partial h$ with the condition
$u^{h}_\eta |_{h=0}= 0$, $(1 \leq \eta \leq m)$,
we obtain $(u_{\eta}^h)_{\eta=1}^{m}$ as a function of $h$.
Similarly, (\ref{Gdiffexperiment}) leads to the expression
of $\partial G^F(h)/\partial h$ in terms of
$(u_{\eta}^h)_{\eta=1}^m$.
The integration yields $G^F(h)$,
where we have used $G^F(h)|_{h=0}=0$.

\begin{figure}[tbh]
\includegraphics[width=8cm]{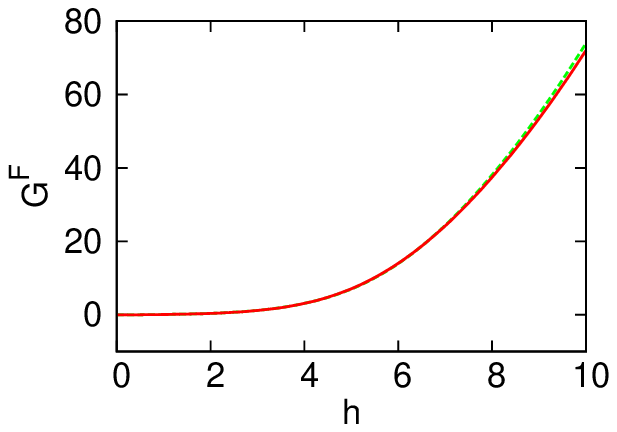}
\includegraphics[width=8cm]{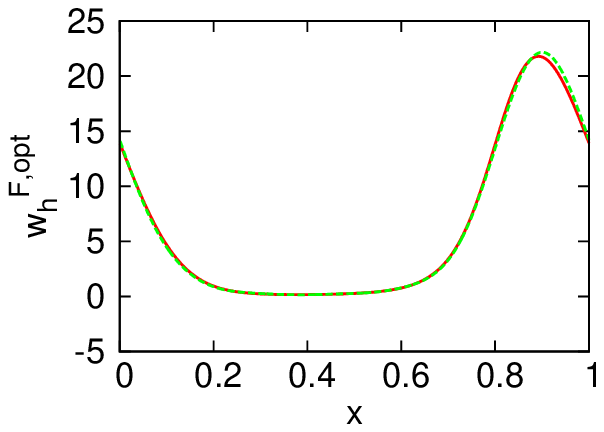}
\caption{(Color online) Numerical experiment for measurement of
$G^F(h)$ (up) and $w_{h}^{F,\rm opt}(x)$ with $h=3$ (down). 
Quantities are
converted to dimensionless forms by setting 
$\gamma=T=L=1$. We fix $f=1$ and $U_0 =3$. 
We assumed to know $T=\gamma=1$ and experimentally determined
$G^F(h)$ and $w_h^{F,\rm opt}(x)$ from trajectories $(x(t))_{t=0}^\tau$
following the method described
in the text. We set $m=10$, $C=10$ and $\tau=400000$.
By taking $10$ samples, we estimated $G^F(h)$ and $w_{h}^{F,\rm opt}(x)$.
The obtained results are displayed with green dashed lines.
Error-bars are within the lines.
The red lines were obtained from the evaluation of the largest eigenvalue 
of $\mathcal L _{ h}^{( x)\dagger}$.
}
\label{figexperiment}
\end{figure}

We performed a numerical experiment of the driven Brownian particle on a ring
subject to a periodic potential $U(x) = U_0\cos(2\pi x/L)$. We set
$\phi_k(x) = \cos(2\pi k x/L)$ $(0\leq k \leq m/2)$, $\phi_k(x) = \sin \left (2\pi x (k-m/2)/L \right )$ 
$(m/2<k\leq m)$. The $m+1$ external forces $(w_k(x))_{k=0}^{m}$
are chosen as $\left ( C\phi_k(x) \right )_{k=0}^{m}$, where $C$ is a constant. Following the recipe we described above, we obtain $G^F(h)$ and
$w^{F,\rm opt}_h(x)$ from trajectories of the particle.
The results are displayed in Fig. \ref{figexperiment}.
We also calculate $G^F(h)$ and $w^{F,\rm opt}_h(x)$ by
approximately evaluating the largest eigenvalue and the
corresponding eigenfunction
of the operator $\mathcal L _{ h}^{( x)\dagger}$.
See Appendix \ref{appendix_largesteigenvalue} for the
calculation method.
These two results are close to each other.
It turns out that our experimental determination method is useful to obtain the
accurate value of $G^F(h)$ from measured trajectories.

\section{Concluding remarks}


In this paper, we have presented the new formula 
(\ref{mainclaim_of_the_paper}) for the cumulant generating function 
of time-averaged velocity. The formula enables us
to determine the cumulant generating function in terms of
the expectation value of time-averaged current
in a modified system. This corresponds to
(\ref{G_equilibrium}) in the example of equilibrium
statistical mechanics, and it is also regarded as an extension of
the fluctuation-dissipation relation.
Furthermore, 
we have pointed out that our formula is related to the additivity principle, 
the Donsker-Varadhan formula, and the least dissipation 
principle. Before ending the 
paper, we address problems that should be studied in future.


First, since our formulas are rather formal, 
we should explore further applications of our formulation 
to various  physical problems.
With regard to this aspect, in Appendix \ref{Appendix_simplemodel}, 
we present a simple model in which we can explicitly calculate 
the cumulant generating function and the optimal force.
We hope that this model might help readers to grasp our formulation
and promote the applications.
Furthermore, although our formulation is valid for interacting particle 
systems, useful predictions are not presented yet. 
A key step in the analysis of many-body problems
is to extract slowly varying effective variables from many-body 
distribution functions. 
Here, we may employ some approximations for extracting
effective variables. The result of the approximate
calculation with our formulation might provide a new insight
into properties of current fluctuations because our formulation
is qualitatively new.


Next, we ask the range of the applicability of our formulation.
The first question may be to consider similar formulas in under-damped Langevin 
systems. The formulas that correspond 
to (\ref{cumulantvariational_expression}) and  (\ref{mainclaim_of_the_paper})
can be derived in a straight-forward manner.
This fact is not surprising at all when the formulation of this paper is 
properly understood.
However, it may be
difficult to find a physical interpretation of the formulas. 
For example, it seems that ``the extra force'' 
appearing in the variational formula  cannot be realized 
in laboratory experiments. More importantly, as far as we attempted,
the variational function may not be expressed in terms of the entropy production
rates, which is different from the over-damped case. We do not understand whether 
or not these are fatal in our formulation. We need further 
investigations.


Here, with regard to the range of the applicability of our formulas,
we restrict our concerns to a formal aspect by putting aside physical 
interpretations. Concretely, 
in Markov jump processes, we can derive the formulas that correspond to
(\ref{cumulantvariational_expression}) and  (\ref{mainclaim_of_the_paper}).
As seen in  (\ref{jump2}) and (\ref{jump4}) in Appendix 
\ref{Appendix_Markov_jump}, the mathematical expressions are quite 
simple. We thus expect that there might be physical examples
for which these formulas are useful. Furthermore, since these are 
expressed in terms of escape rates, it is  conjectured that
similar variational principles are formulated in  many-body 
Hamiltonian systems. It is stimulating to prove this conjecture.
By combining physical interpretations and formal aspects, 
we wish to seek for a universal structure behind our formulation.


Finally, we go back to our basic motivation. Our naive
idea is that the large deviation theory for current 
fluctuations might be associated with an operational 
framework. This was conjectured from the well-established
fact that the large deviation theory for fluctuations of thermodynamic
variables
is tightly connected to  equilibrium thermodynamics. 
The progress developed in this paper does not provide 
a final answer to the problem, but the results are
encouraging us to continue our efforts. In the best
case, we might have an operational framework of
thermodynamics for currents. Then, we will consider
a possibility to unify
such a framework with equilibrium thermodynamics.
It may be plausible that a unified scheme is related to the so-called 
steady state thermodynamics \cite{Oono_Paniconi, Hatano-Sasa, Sasa-Tasaki, KNST, KNST_regular, Saito_Tasaki}. With dreaming such possibilities, we 
challenge to study operational viewpoints in non-equilibrium 
statistical mechanics.


\section*{Acknowledgement}

The authors thank H. Tasaki for continuous discussions on 
non-equilibrium statistical mechanics. 
They also thank F. van Wijland for indicating the exactly calculable
model in Appendix \ref{Appendix_simplemodel}.
The present study was 
supported by grants from the Ministry of Education, Culture, 
Sports, Science, and Technology of Japan, Nos. 22340109 and 23654130.

\appendix


\section{Equilibrium statistical mechanics}
\label{appendix_thermodynamics}

We review the relation between the cumulant generating function of a
thermodynamic variable and a thermodynamic function.
Specifically, we consider a spin model with finite range interaction
in a finite
dimensional lattice $\Lambda$. Let $\sigma_i$, $i\in \Lambda$, be a spin variable
whose value is either $+1$ or $-1$. We denote
$\left (\sigma_i \right )_{i\in \Lambda}$ by $\sigma$ collectively.
We assume that the Hamiltonian for the system under
a magnetic field $H$ is given by
\begin{equation}
\mathcal{H}_H(\sigma)=\mathcal{H}_0(\sigma) -H\mu \sum_{i\in \Lambda}\sigma_{i} ,
\label{hamiltonian}
\end{equation}
where $\mu$ is the magnetic moment.
According to equilibrium statistical mechanics, for any
observable $A(\sigma)$, the expectation value $\langle A \rangle_{T,H}$
in the system at temperature $T$ is given by
\begin{equation}
\langle A \rangle _{T ,H} = \sum_{\sigma}A(\sigma)\frac{e^{- \mathcal H_H(\sigma)/T}}
{Z(T, H)},
\end{equation}
where 
\begin{equation}
Z(T,H)=\sum_{\sigma}e^{- \mathcal H_H(\sigma)/T}.
\end{equation}
As an observable, we consider the magnetization density
\begin{equation}
\mathtt m (\sigma)=\frac{\mu}{N}\sum _{i\in \Lambda}\sigma_i,
\end{equation}
where $N=|\Lambda|$.
Now, we define the free energy density $f(T,H)$ by
\begin{equation}
f(T, H) =- T \lim _{N \rightarrow \infty}\frac{1}{N} \log Z(T,H) \label{gibbs_free_energy}.
\end{equation}
Then, the fundamental relation of thermodynamics is written as
\begin{equation}
df=-sdT-mdH, \label{thermodynamic_relation}
\end{equation}
where $s$ is the entropy density and $m$ is the expectation value of the
magnetization density $\mathtt m(\sigma)$. Notice that
$f(T,H)$ is determined from the measurement of the heat capacity
$c(T,H)$ and $m(T,H)$.

Putting aside the thermodynamic function, we consider the scaled cumulant 
generating function of $\mathtt m(\sigma)$, which is defined as
\begin{equation}
G_{T,H}(h) = 
\lim _{N \rightarrow \infty}\frac{1}{N} \log \langle e^{h\mathtt m(\sigma)N}\rangle _
{T, H}  \label{zikacumulant}.
\end{equation}
From (\ref{gibbs_free_energy}) and (\ref{zikacumulant}), we find the following relation between
the cumulant generating function and the free energy:
\begin{equation}
G_{T,H}(h) = -\frac{1}{T} \left [  f\left (T,H+ hT \right ) - f(T,H) \right ] \label{cumu_vs_gibbs}.
\end{equation}
In particular, from the thermodynamic relation (\ref{thermodynamic_relation}),
we obtain
\begin{equation}
\frac{\partial G_{T,H}(h)}{\partial h} 
= m(T, H+hT)
\label{G_equilibrium}
\end{equation}
for any $h$.
By differentiating (\ref{G_equilibrium}) with respect to $h$ and set $h=0$,
we obtain 
\begin{equation}
\lim_{N\rightarrow \infty} N\left \langle \left (\mathtt m(\sigma) - m \right ) ^2\right \rangle_{T,H}=
T \frac{\partial m(T,H)}{\partial H},
\end{equation}
which is referred to as a fluctuation-response relation \cite{callen}.


\section{Properties of the largest eigenvalue}
\label{Appendix_largest_eigenvalue}

We show that the largest eigenvalue of
the operators ${\mathcal L}_{\bv h}^{(\bv x)}$ and 
${\mathcal L^{\dagger}}_{\bv h}^{(\bv x)}$ is real and that the corresponding 
eigenfunctions are
positive. 
The argument presented here 
is not mathematically rigorous, but provides a practical idea for understanding the 
claim.
Although we focus on one-dimensional case for simplicity, the extension to 
higher-dimensional cases are straight-forward.
We express the operator $\mathcal L_h^{(x)}$ in the form
\begin{equation}
{\mathcal L}_h ^{(x)}\cdot \varphi =A(x)\frac{\partial^2}{\partial x^2}\varphi + 
B(x) \frac{\partial}{\partial x}\varphi +   C(x)\varphi  ,
\end{equation}
where $A(x)>0$.
We study the eigenvalue problem of $\mathcal L_h^{(x)}$
in the space of periodic functions defined on the interval $[0, L]$.

We define $x_i\equiv Li/N$, $i=1,2,...,N$, and let $\Delta x$ be the distance
between $x_{i+1}-x_i$. 
We replace the differential operator $\partial /\partial x$ and 
$\partial^2 /\partial x^2$ by differences.
Concretely, the following replacements are considered:
\begin{equation}
\frac{\partial \varphi}{\partial x}\Big{|}_{x=x_i}  \rightarrow 
\frac{\varphi( x_{i+1} )-\varphi( x_{i-1})}{2 \Delta x},
\end{equation}
\begin{equation}
\frac{\partial ^2\varphi}{\partial x^2} \Big{|}_{x=x_i} \rightarrow \frac{ \varphi(x_{i+1})+
\varphi(x_{i-1})- 2\varphi(x_i)}{(\Delta x)^2},
\end{equation}
for $i=2,3,...,N-1$. A special care is needed for the cases $i=1$ and $i=N$.
For example, 
\begin{equation}
\frac{\partial \varphi}{\partial x}\Big{|}_{x=x_1}  \rightarrow 
\frac{\varphi( x_{2})-\varphi( x_{N})}{2 \Delta x},
\end{equation}
\begin{equation}
\frac{\partial ^2\varphi}{\partial x^2} \Big{|}_{x=x_N} \rightarrow \frac{ \varphi
(x_{1})+
\varphi(x_{N-1})- 2\varphi(x_N)}{(\Delta x)^2}.
\end{equation}
Then, we define an $N \times N$ matrix $M_N$ such that
\begin{equation}
\begin{split}
&(M_N {\bv \varphi})_i  \\
& =A(x_i)\frac{\partial ^2\varphi}{\partial x^2} \Big{|}_{x=x_i}  + 
B(x_i) \frac{\partial \varphi}{\partial x}\Big{|}_{x=x_i} +  C(x_i)\varphi(x_i),
\end{split}
\end{equation}
where the $N$ dimensional vector $\bv{\varphi}$
is defined by $({\bv \varphi})_i = \varphi(x_i)$.
Since $M_N$ is an approximation of the operator $\mathcal L_h^{(x)}$,
we assume that the eigenvalues and eigenfunctions of $\mathcal L_h^{(x)}$
are determined from the limit $N\rightarrow \infty$ for the eigenvalues and 
eigenvectors of $M_N$.
Thus, below, we study the eigenvalue problem of $M_N $ instead of the eigenvalue problem of $\mathcal L_h^{(x)}$.

We show that the largest eigenvalue of $M_N$ is real and that all components of the corresponding left and right eigenvectors are positive, when $N$ is sufficiently
large.
This claim is understood from the Perron-Frobenius theory.
A real matrix $B$ is called an irreducible ML-matrix \cite{Seneta} if $B$ satisfies 
the following
two conditions: (i) All non-diagonal components of $B$ are non-negative. (ii) There
exists a positive number $a$ such that the matrix $T\equiv a I + B$ is
irreducible, where $I$ is the unit matrix. 
The Perron-Frobenius theory leads to the statement that
the irreducible ML-matrix $B$ has an eigenvalue $\mu_{\rm pf}$ that
satisfies the following conditions:
\begin{enumerate}
\item $\mu_{\rm pf}$ is real.
\item  The associated left and right eigenvectors
are strictly positive and unique up to constant multiples.
\item  For any eigenvalue $\mu$ except for $\mu_{\rm pf}$,
the real part of $\mu$ is less than $\mu_{\rm pf}$.
\end{enumerate}
See Theorem 2.6 of Ref. \cite{Seneta}. 
Since $A(x)>0$, the matrix $M_N$ for sufficiently large $N$ 
turns out to be an irreducible ML-matrix. We thus obtain the claim
that the largest eigenvalue of $M_N$ is real and that all components of the corresponding left and right eigenvectors are positive.


\section{$\sigma ^{\bv F}(P)$ and entropy production rate}
\label{appendix_entropyproduction}
We explain thermodynamic interpretations of the first and second terms
of (\ref{psi_expression}), which were denoted by $\Sigma^{\bv F}(P)$
and $\Omega^{\bv F}(P)$ in Section \ref{defference_of_entropy_production}, 
respectively.

We first consider the first term, $\Sigma^{\bv F}(P)$.
We assume that the system is set under the force $\bv F$ and that the probability
density is given by $P$. We define a probability density 
$p^{\bv F}(\bv x,t)$ as the solution of
\begin{equation}
\frac{\partial}{\partial t}p^{\bv F}(\bv x,t) = - \sum_{i=1}^{Nd}\frac{\partial}{\partial x_i} 
q_i^{\bv F}(\bv x|p^{\bv F})
\label{time_derivative_P}
\end{equation}
with the initial condition
\begin{equation}
p^{\bv F}(\bv x , 0) = P(\bv x).
\label{initial_condition_P}
\end{equation}
Here, we take the limit $t\rightarrow 0$ in (\ref{time_derivative_P}).
By noting (\ref{initial_condition_P}) in the right-hand side of (\ref{time_derivative_P}), we obtain
\begin{equation}
\lim_{t \rightarrow 0}\frac{\partial}{\partial t} p^{\bv F}(\bv x,t) =
- \sum_{i=1}^{Nd}\frac{\partial}{\partial x_i} q_i^{\bv F}(\bv x |P).
\label{initial_derivative}
\end{equation}
Thus, the first term of
(\ref{psi_expression}), $\Sigma^{\bv F}(P)$ is expressed as
\begin{equation}
\Sigma^{\bv F}(P)=-\lim_{t\rightarrow 0}\left \{ \int _{\Omega ^N} d\bv x \left [ \log p^{\bv F}(\bv x,t)
 \right ] \frac{\partial}{\partial t} p^{\bv F}(\bv x,t) \right \} .
\label{appendix_S}
\end{equation}
Since $\int _{\Omega ^N} d\bv x p^{\bv F}(\bv x,t)=1$, (\ref{appendix_S})
is further rewritten as
\begin{equation}
\Sigma^{\bv F}(P)=  \frac{d}{d t}\left (-\int _{\Omega ^N} d\bv x p^{\bv F}(\bv x,t)  \log p^{\bv F}(\bv x,t)
\right ) \bigg{|}_{t=0} .
\label{S_expression}
\end{equation}
This shows that $\Sigma ^{\bv F}(P)$ is the increasing rate of the Shannon entropy
for a given distribution $P$.

Next, we consider the second term, $\Omega ^{\bv F}(P)$. 
We shall show that $\Omega ^{\bv F}(P)$ is the expectation value of the
energy dissipation rate divided by $T$
when the distribution is given by $P$.
According to the energetic interpretation of Langevin equations \cite{Sekimoto, Stochastic_Energetics},
the energy dissipation rate $\Psi$ for each realization is given by
\begin{equation}
\Psi = \sum_{i=1}^{Nd}\left ( \gamma _{i}  \dot x_{i} - \sqrt{2\gamma _i T} \xi_i  \right )
 \circ \dot x_i,
 \label{dissipation_rate_sekimoto}
\end{equation}
where the symbol $\circ$ represents the multiplication in the sense of
the Stratonovich \cite{Gardiner}.
It should be noted that $ \Psi$ is a stochastic variable. By substituting
(\ref{langevin}) into (\ref{dissipation_rate_sekimoto}),
we write
\begin{equation}
\Psi = \sum_{i=1}^{Nd}\left ( \frac{1}{\gamma_i}F_i(\bv x)^2 + 
\sqrt{\frac{2T}{\gamma_i}}F_i(\bv x)\circ \xi_i \right ).
\label{dissipation_rate_sekimoto2}
\end{equation}
Here, we change the multiplication rule from the Stratonovich type ($\circ$) to 
the Ito type ($\cdot$) \cite{Gardiner}. Then, (\ref{dissipation_rate_sekimoto2})
 becomes
\begin{equation}
\Psi = \sum_{i=1}^{Nd}\left ( \frac{1}{\gamma_i}F_i(\bv x)^2 + \frac{T}{\gamma_i}
\frac{\partial F_i(\bv x)}{\partial x_i} +
\sqrt{\frac{2T}{\gamma_i}}F_i(\bv x)\cdot \xi_i \right ).
\label{appendix_psihat}
\end{equation}
By taking the expectation value of (\ref{appendix_psihat}) and integrating by parts
the second term, we obtain
\begin{equation}
\begin{split}
\left \langle  \Psi \right \rangle &=\int d\bv  x
 \sum_{i=1}^{Nd}F_i(\bv x) \left ( \frac{1}{\gamma_i}F_i(\bv x) P(\bv x)
  - \frac{T}{\gamma_i}
\frac{\partial  P(\bv x)}{\partial x_i} \right ) \\
&=T\Omega^{\bv F}(P), \label{heat_rate}
\end{split}
\end{equation}
where we have used $\left \langle F_i(\bv x)\cdot \xi_i \right \rangle=0$.
The last expression claims that 
$\Omega ^{\bv F}(P)$ is the expectation value of the
energy dissipation rate divided by $T$
when the distribution is given by $P$.


\section{Iterative expression of cumulant coefficients}
\label{appendix_iteratibe_expression}

For the system studied in Section \ref{Section_application},
we determine the cumulant
coefficients in an iterative manner.
We start with the expression
\begin{equation}
2TG^F(h) = \frac{T}{\gamma} \frac{\partial u^F_h(x)}{\partial x}  + 
\frac{u^F_h(x)}{\gamma}
\left ( F(x) + \frac{u^F_h(x)}{2} \right ), \label{cumuderivation1}
\end{equation}
which is obtained from (\ref{cumulant}) with $d=N=1$,
where $u_h^F(x)$ satisfies
\begin{equation}
\int _{0}^{L}dx u^F_h(x) = 2ThL,\label{cumuderivation2}
\end{equation}
\begin{equation}
u^F_h(0) = u^F_h(L). \label{cumuderivation3}
\end{equation}
From a trivial relation
\begin{equation}
\begin{split}
& \frac{\partial u^F_h(x)}{\partial x} 
+ \frac{1}{T} u^F_h(x) F(x)  \\
&= e^{ -  \int _{0}^{x}dyF(y)/T } 
\frac{\partial}{\partial x} \left [ u^F_h(x) 
e^{ \int _{0}^{x}dyF(y)/T } \right ],
\end{split}
\end{equation}
we rewrite (\ref{cumuderivation1}) as
\begin{equation}
\begin{split}
& \frac{\partial}{\partial x} \left [ u^F_h(x) 
e^{  \int _{0}^{x}dyF(y)/T}\right ] \\
& = 
e^{ \int _{0}^{x}dyF(y)/T } \left ( 2\gamma G^F(h) - \frac{1}{2T}u^F_h(x)^2 \right ).
\label{iterative1}
\end{split}
\end{equation}
By integrating (\ref{iterative1}) with respect to $x$, we obtain
\begin{equation}
\begin{split}
&u^F_h(x) = e^{- \int _{0}^{x}dyF(y)/T}  \\
& \times \left [C + \int _{0}^{x}dy 
 e^{ \int _{0}^{y}dzF(z)/T} \left ( 2\gamma G^F(h) - \frac{1}{2T}u_{h}^F(y)^2 \right )   \right ], \label{cumuderivation4}
\end{split}
\end{equation}
where $C$ is an integration constant.
Here, by direct calculation, we can derive an identity
\begin{equation}
\begin{split}
&\int^{x}_{0} dy g(y) e^{ \int_{0}^{y} dzF(z)/T} = \frac{1}{1-e^{ f L/T}} \\
&\left [ \int _{0}^{L} dy g(y) e^{ \int _{0}^{y}dzF(z)/T} - \int _{0}^{L}dy
g(x+y)e^{ \int _{0}^{x+y}dzF(z)/T} \right ] \label{cumuderivation6}
\end{split}
\end{equation}
for any periodic function $g(x)$.
We then set
\begin{equation}
g(x)=2\gamma G^F(h) - \frac{1}{2T}u_{h}^F(x)^2 
\end{equation} 
in (\ref{cumuderivation6}) and substitute it into 
the second term of the right-hand side of (\ref{cumuderivation4}). 
The result is
\begin{equation}
\begin{split}
& u^F_h(x) =  e^{- \int _{0}^{x}dyF(y)/T} \left[ \tilde C  
 - \frac{1}{1-e^{ fL/T}} \right. \\
 &\times \left. 
\int _{0}^{L}dy 
 e^{ \int _{0}^{x+y}dzF(z)/T} \left ( 2\gamma G^F(h) - \frac{1}{2T}
 u_h^F(x+y)^2 \right ) 
\right],
\label{D9}
\end{split}
\end{equation}
where $\tilde C$ is a constant. 
The condition $u^F_h(0)=u^F_h(L)$ leads to 
$\tilde C =0$. Through the transformation of the integration variable 
$y\rightarrow L-y$, we rewrite (\ref{D9}) as
\begin{equation}
\begin{split}
& u^F_h(x) =\frac{2\gamma}{1-e^{- f L/T}} \\
&\left [ G^F(h) I_{+}(x) - 
\frac{1}{4T\gamma} \int _{0}^{L}dy 
 e^{ \int _{x}^{x-y}dzF(z)/T} 
 u^F_h(x-y)^2\right ] , 
\label{cumuderivation7}
\end{split}
\end{equation}
where we have used (\ref{Iplusminus}).
By integrating the both sides of (\ref{cumuderivation7}) with 
respect to $x$ and using the condition (\ref{cumuderivation2}),
we derive 
\begin{equation}
G^F(h) = hL \frac{T}{\gamma}(1-e^{- f L/T}) \alpha  + \frac{\alpha }{4T\gamma}
\int _{0}^{L}dx I_{-}(x) u_h^F(x)^2, \label{G(h)iteration}
\end{equation}
where $\alpha$ was defined in (\ref{Ipm_alpha}).

It should be noted that $u_h^F(x)$ satisfies a self-consistent equation. 
Indeed,
by substituting (\ref{G(h)iteration}) into (\ref{cumuderivation7}), we find
\begin{equation}
\begin{split}
& u_h^F(x) =2ThL \alpha I_{+}(x) + \frac{1 }{2T(1-e^{- fL/T})} 
 \int _{0}^{L}dy \\
& \left [  e^{ \int _{x}^{x-y}dzF(z)/T} \left ( \alpha \int _{0}^{L}dr I_{-}(r)u_h^F(r)^2 - u_h^F(x-y)^2 \right ) \right ] . \label{u(x)iteration}
\end{split}
\end{equation}
By substituting (\ref{remarks_reimann_3}) into (\ref{u(x)iteration}),
we can determine the coefficient $u_n(x)$ iteratively.
Concretely, by noting
\begin{equation}
\begin{split}
u^F_h(x)^2 & = \sum_{n=1}^{\infty}\sum_{l=1}^{\infty}u_n(x)u_l(x)h^{n+l} \\
&= \sum_{n=2}^{\infty} h^n \left ( \sum_{l=1}^{n-1}u_l(x)u_{n-l}(x) \right ),
\label{u_hsquare}
\end{split}
\end{equation}
we derive a recursion equation
\begin{equation}
\begin{split}
&u_n(x) =\frac{1}{2T(1-e^{- f L/T})} \sum_{l=1}^{n-1} \int _{0}^{L}dy 
 e^{ \int _{x}^{x-y}dzF(z)/T} \\
 &\left [ \alpha \left(  \int _{0}^{L}dr I_{-}(r)u_l(r)u_{n-l}(r) \right )
 - u_l(x-y)u_{n-l}(x-y) \right ] 
\end{split} 
\end{equation}
for $n \geq 2$.
For example, by recalling (\ref{remark_reimann_5}), we calculate
\begin{equation}
\begin{split}
 u_2(x)  =&\frac{2T\alpha^2L^2}{(1-e^{- f L/T})} \int _{0}^{L}dy 
 e^{ \int _{x}^{x-y}dzF(z)/T} \\
& \left [ \alpha \left ( \int _{0}^{L}dr I_{-}(r)I_+(r)^2 \right )
 - I_+(x-y)^2 \right ] .
 \end{split}
\end{equation}

Furthermore, we expand $G^F(h)$ as
\begin{equation}
G^F(h)=\sum_{n=1}^{\infty} \frac{G_n}{n!} h^n.
\label{G_hexpansion}
\end{equation}
By substituting (\ref{u_hsquare}) and (\ref{G_hexpansion})
into (\ref{G(h)iteration}), we obtain
\begin{equation}
G_1=L\frac{T}{\gamma}(1-e^{- f L/T})\alpha
\label{G_1iterative}
\end{equation}
and
\begin{equation}
G_n=\sum_{l=1}^{n-1}\frac{\alpha n! }{4T\gamma} \int _{0}^{L}dxI_{-}(x)u_{l}(x)
u_{n-l}(x)
\label{G_niterative}
\end{equation}
for $n\geq 2$. With this formula, the cumulant coefficients $G_n$ can be calculated
in an iterative manner.
For example, the third-order cumulant coefficient is derived as
\begin{equation}
G_3=\frac{6\alpha^2 L}{\gamma}\int _{0}^{L}dx I_-(x)I_+(x)u_2(x).
\end{equation}

 Reimann {\it et al} found the exact expression of the diffusion constant
(\ref{remark_reimann_9}) from the first and second cumulants
of first passage time \cite{Reimann}. 
Here, we remember an iterative formula for the moments of the first passage time
\cite{Gardiner}. By comparing 
 this iterative formula with our iterative formula, one might find
a general relationship between the cumulant generating function
and statistical properties of the first passage time beyond the first two cumulants.


\section{
Calculation of $G(h)$ and $w_h^{\rm opt}(x)$}
\label{appendix_largesteigenvalue}

We focus on the case that $N=1$ and $d=1$, which is studied in section
\ref{Section_application}.
We numerically solve the largest eigenvalue problem (\ref{eigenad}).
First, we consider a linear equation
\begin{equation}
\frac{\partial }{\partial t}\phi(x,t) = {\mathcal L}^{(x)\dagger}_{h} \phi(x,t).
\label{phi_zikannhattenn}
\end{equation}
Since $G^F(h)$ is the largest eigenvalue of $\mathcal L_{h}^{(x)\dagger}$,
the long time behaviors of solutions are described as
\begin{equation}
\phi (x,t)\simeq  \psi_h(x) e^{G^F(h)t}
\label{phi_zennkin}.
\end{equation}
Now, we assume $\phi(x,t)>0$ for any $t$ by choosing appropriate initial
conditions. Then, we define
\begin{equation}
u(x,t) \equiv 2T \frac{\partial }{\partial x} \log \phi(x,t) +2Th \label{u(x,t)def}.
\end{equation}
By dividing (\ref{phi_zikannhattenn}) by $\phi(x,t)$ and using the same
method deriving (\ref{cumulant}) in section III.A, we obtain
\begin{equation}
2T\frac{\partial}{\partial t} \log \phi(x,t) 
=\frac{F(x)}{\gamma}u(x,t)
+\frac{1}{2\gamma}u(x,t)^2
+\frac{T}{\gamma}\frac{\partial}{\partial x} 
u(x,t).
\label{logphi_timeevolution}
\end{equation}
Finally, by differentiating (\ref{logphi_timeevolution}) with respect to $x$,
we rewrite (\ref{logphi_timeevolution}) as
\begin{equation}
\frac{\partial}{\partial t} u(x,t) = 
-\frac{\partial}{\partial x} j_{u}(x,t)\label{experiment_continue},
\end{equation}
with a current
\begin{equation}
j_u(x,t) \equiv- \frac{F(x)}{\gamma}u(x,t)
-\frac{1}{2\gamma}u(x,t)^2
-\frac{T}{\gamma}\frac{\partial}{\partial x} u(x,t) \label{experiment_current}.
\end{equation}
Let $u_*(x)$ be the stationary solution given as the long-time limit of $u(x,t)$.
Then, from (\ref{phi_zennkin}) and (\ref{u(x,t)def}), we obtain
\begin{equation}
u_*(x)=2Th\frac{\partial}{ \partial x}\log \psi_h(x)+2Th.
\end{equation}
By recalling (\ref{wopt_psi}),
we have
\begin{equation}
w_h^{\rm opt}(x) = u_*(x).
\end{equation}
Furthermore, by comparing (\ref{experiment_current})
with (\ref{cumulant}), we obtain
\begin{equation}
G^F (h)= -\frac{1}{2T}\lim_{t \rightarrow \infty} j_{u}(x,t) .
\end{equation}
Therefore, by numerically solving (\ref{experiment_continue}),
we can calculate $G^F(h)$ and $w_h^{\rm opt}(x)$.



\section{Exactly calculable model}
\label{Appendix_simplemodel}

We illustrate our theoretical idea by explicitly deriving 
the cumulant generating function and the optimal force
for  a linear Langevin equation. The model
describes the motion of a single Brownian particle
in two dimensions, which corresponds to the case that 
$N=1$ and $d=2$ in (\ref{langevin}). By setting $\gamma=1$ 
for notational simplicity,  we write the equation as
\begin{equation}
\frac{d\bv x}{dt}=\bv F(\bv x)+\sqrt{2T}\bv \xi,
\end{equation}
with 
\begin{equation}
\bv F(\bv x)\equiv -k \bv x + \epsilon \bv l(\bv x),
\end{equation}
where $\bv l =(-x_2,x_1)$.
From (\ref{stationarryrho}) and (\ref{stationarrycontinue}), 
we  obtain  the stationary distribution as
\begin{equation}
P_{\rm st}^{\bv F}(\bv x) =C 
\exp\left (-\frac{k}{2T}\bv x^2 \right ),
\label{stationarydist}
\end{equation}
with a normalization constant $C$. 
Although the stationary distribution is Gaussian,
the detailed balance condition is not satisfied because of 
the circulation force $\epsilon \bv l$.
One can check this fact immediately
from (\ref{stationarryrho}) and (\ref{stationarydist}).
It  should be noted that the linear Langevin equation
was studied with a novel idea, irreversible circulation,
which characterizes the violation of detailed balance
\cite{Tomita-Tomita}.

The quantity we consider here is a time-averaged 
angular momentum (or irreversible circulation)
defined as
\begin{equation}
L(\tau)\equiv \frac{1}{\tau}\int_{0}^{\tau} 
dt \bv l(\bv x(t)) \cdot {\dot { \bv x}(t)}.
\label{heatdef}
\end{equation}
For the scaled cumulant generating function of $L$, 
\begin{equation}
G(h)=\lim_{\tau\rightarrow \infty} \frac{1}{\tau}\log \left \langle 
e^{\tau h L(\tau)}\right \rangle,
\end{equation}
we can derive the variational principle 
by  using the same method as Section \ref{Subsection_derivation}. 
First, we define an operator $\mathcal L_{h}$ as 
\begin{equation}
\begin{split}
\mathcal L_{h} \cdot \varphi &  = T \bv \nabla ^2 \varphi 
- \bv \nabla\cdot \left (\bv F \varphi \right )
+ h \left ( \bv l \cdot \bv F  \right ) \varphi  \\
&+ h^2T\bv  l^2 \varphi
-2Th\bv \nabla \cdot \left ( \bv l \varphi \right ).
\end{split}
\end{equation}
Since $G(h)$
is equal to the largest eigenvalue of the operators 
$\mathcal L_{h}$ and $\mathcal L_{h}^{\dagger}$, 
we have 
\begin{equation}
\mathcal L_{h}^{\dagger}\cdot \phi = G(h) \phi,
\label{eigenproblem}
\end{equation}
where $\phi$ is the positive eigenfunction 
corresponding to the largest eigenvalue of $\mathcal L_{h}^{\dagger}$.
Next, we define a force $\bv u$ as
\begin{equation}
\bv u(\bv x) = 2T\left (\bv \nabla \log \phi(\bv x) + h \bv l(\bv x) \right ).
\end{equation}
By dividing (\ref{eigenproblem}) by $\phi$ and using the definition
of $\bv u$, we obtain
\begin{equation}
\bv F \cdot \bv u + \frac{1}{2} \bv u ^2 + T\bv \nabla  \cdot \bv u = 2T G(h).
\label{aftercolehopf}
\end{equation}
This takes the same form as (\ref{cumulant}) so that
we can follow the argument from (\ref{cumulant}) to (\ref{G(h)experiment}).
We then arrive at 
\begin{equation}
G(h)=\tilde \Phi_{h}(\bv w) + \frac{1}{4T}\left \langle 
(\bv w - \bv u)^2\right \rangle _{\rm st}^{\bv F+\bv w},
\label{key}
\end{equation}
with
\begin{equation}
\tilde \Phi_{h}(\bv w) 
=h \int_{\Omega} d\bv x \bv l\cdot \bv j_{\rm st}^{\bv F+\bv w}
-\frac{1}{4T} \left \langle\bv w^2 \right \rangle_{\rm st}^{\bv F+\bv w}.
\label{phi}
\end{equation}
Here, it should be noted that 
\begin{equation}
\int_{\Omega} d\bv x \bv l\cdot \bv j_{\rm st}^{\bv F+\bv w}
=\left \langle L \right \rangle_{\rm st}^{\bv F+\bv w}.
\end{equation}
(\ref{key}) is the key identity of our formulation, which corresponds to (\ref{G(h)experiment}).
Indeed, from (\ref{key}) we obtain
\begin{equation}
G(h)=\max_{\bv w} \Phi_{h}(\bv w),
\label{Gvar}
\end{equation}
\begin{equation}
\frac{\partial G(h)}{\partial h}= \left \langle L \right \rangle_{\rm st}^{\bv F+\bv w_{\rm opt}},
\end{equation}
where the optimal force $\bv w_{\rm opt}$ is given as
\begin{equation}
\bv w_{\rm opt}(\bv x) =\bv u(\bv x).
\end{equation}

Finally, by solving (\ref{aftercolehopf}), we derive the explicit 
expression of $\bv u$ and $G(h)$. Suppose that 
\begin{equation}
\phi = \exp\left ( -\frac{a}{2}\bv x ^2 \right ),
\label{F18}
\end{equation}
where $a$ is a constant.
This yields 
\begin{equation}
\bv u(\bv x) = -2Ta\bv x +2Th \bv l(\bv x).
\label{F19}
\end{equation}
By substituting (\ref{F19}) into (\ref{aftercolehopf}),
we obtain
\begin{equation}
G(h)=-2Ta
+ \bv x^2\left [T a^2 +k a +h(Th +\epsilon) \right ].
\end{equation}
Since the left-hand side is independent of $\bv x$,
we determine the constant $a$ as
\begin{equation}
a=\frac{1}{2T} \left (-k +\sqrt{k^2-4Th(Th+\epsilon)
} \right ).
\label{optimal_a}
\end{equation}
Here, we can confirm that (\ref{eigenproblem}) holds for (\ref{F18})
with (\ref{optimal_a}), because (\ref{eigenproblem})
is equivalent to (\ref{aftercolehopf}). We then obtain
\begin{equation}
G(h) = k-\sqrt{k^2 -4Th(Th+\epsilon)}.
\label{exactG}
\end{equation}
From (\ref{F19}) and (\ref{optimal_a}), the force in the optimal system is explicitly written as
\begin{equation}
\bv F(\bv x) +\bv u(\bv x)
=-\sqrt{k^2-4Th(Th+\epsilon)}\bv x + (\epsilon+2Th) \bv l(\bv x). 
\end{equation}


\section{Formula in Markov jump process}
\label{Appendix_Markov_jump}


The model we consider here is a continuous-time Markov process
of a finite space $\mathcal M$. We assume that the transition rate, 
$w(\bv a \rightarrow \bv a^{\prime})$ for $\bv{a} \not =   \bv{a^{\prime}}
\in {\mathcal M}$, is irreducible. We set 
$w(\bv a \rightarrow \bv a)=0$ for notational simplicity.
The escape rate $\lambda (\bv a)$ is then defined  by
\begin{equation}
\lambda (\bv a) 
= \sum_{\bv a^{\prime} \in \mathcal M} w(\bv a \rightarrow \bv a^{\prime}).
\end{equation}
Let  $P^w(\bv a,t)$ be a probability of $\bv a(t)=\bv a$, where 
the superscript $w$ represents the transition rate dependence 
of the probability $P^w$.
Then, $P^w(\bv a,t)$ satisfies the master equation
\begin{equation}
\frac{\partial}{\partial t}P^{w}(\bv a, t) 
= \sum_{\bv a^{\prime}} 
[w(\bv a^{\prime} \rightarrow \bv a)- \lambda(\bv a)\delta_{\bv a,\bv a^{\prime}}]
P^{w}(\bv a^{\prime}, t).
\end{equation}
For later convenience, we define a matrix by 
\begin{equation}
\mathcal A(\bv a,\bv a^{\prime}) 
\equiv 
w(\bv a^{\prime} \rightarrow \bv a)- \lambda(\bv a)\delta_{\bv a,\bv a^{\prime}}.
\end{equation}
The stationary distribution $P^w_{\rm st}(\bv a)$ satisfies
\begin{equation}
\sum_{\bv a^{\prime}} \mathcal A(\bv a,\bv a^{\prime}) P_{\rm st}^{w}(\bv a^{\prime}) =0.
\label{jump_stationary}
\end{equation} 
We denote by $\left \langle g \right \rangle_{\rm st}^{w}$
the expectation value of a function $g(\bv a)$ with respect to the stationary
distribution $P_{\rm st}^{w}(\bv a)$.


We fix a time interval $[0,\tau]$.  For a given history 
$(\bv a(t))_{t=0}^{\tau}$, where $\bv a(t) \in {\mathcal M}$,
a sequence of transition times $(t_1,t_2,\cdots,t_{n-1})$ is 
determined, and $\bv a (t)$ is expressed as 
$\bv a (t)=\bv a_i$  for  $ t_{i-1} < t < t_{i}$ with
$t_0=0$ and $t_n=\tau$. Here, we consider a quantity 
$\alpha(\bv a_{i},\bv a_{i+1})$ defined for the jump event 
$\bv a_{i} \to \bv a_{i+1}$ at $t=t_i$ ($1\leq i \leq n-1$). 
Its time-averaged value in the history is 
written as 
\begin{equation}
A(\tau) = \frac{1}{\tau} 
\sum_{i=1}^{n-1}\alpha(\bv a_{i},\bv a_{i+1}).
\end{equation}
We then study  the scaled cumulant generating function of 
$A(\tau)$, which is defined by
\begin{equation}
G(h)\equiv \lim_{\tau \rightarrow \infty}\frac{1}{\tau} 
\log \left \langle e^{\tau h A(\tau)} \right \rangle,
\end{equation}
where $\bra \ \ket$ represents the sample average
over realizations of history with an initial distribution.


It is known that $G(h)$ is equal to the largest eigenvalue
of the matrix $\mathcal A_h$ given by
 \begin{equation}
\mathcal A_h(\bv a^{\prime},\bv a) 
= w(\bv a \rightarrow \bv a^{\prime})
e^{h\alpha(\bv a,\bv a^{\prime})} - \lambda(\bv a)\delta_{\bv a,\bv a^{\prime}}.
\end{equation}
See Ref. \cite{garrahan} for the derivation.
Let $\phi_{\rm max}(\bv a)$ be the left eigenvector for the largest 
eigenvalue of $\mathcal A_h$. That is,
\begin{equation}
\sum_{\bv a^{\prime}} \phi_{\rm max} (\bv a^{\prime}) \mathcal A_h( \bv a^{\prime},\bv a)
= G(h) \phi_{\rm max}(\bv a).
\label{eigenvalue_equation_jump}
\end{equation}
Since $\phi_{\rm max}(\bv a) > 0$ 
(See Appendix \ref{Appendix_largest_eigenvalue}), 
we may define a potential function $V_{\rm max}(\bv a)$ as
\begin{equation}
V_{\rm max}(\bv a) \equiv -2 \log \phi_{\rm max}(\bv a) . 
\end{equation}
Dividing the both sides of (\ref{eigenvalue_equation_jump}) by 
$\phi_{\rm max}(\bv a)$, we obtain
\begin{equation}
\begin{split}
&G(h)\\
&=
\sum_{\bv a^{\prime}}\Big{[} 
 w(\bv a \rightarrow \bv a^{\prime})e^{-V_{\rm max}(\bv a^{\prime})/2
+V_{\rm max}(\bv a)/2 + h\alpha(\bv a,\bv a^{\prime})}   \\
& \qquad - w(\bv a \rightarrow \bv a^{\prime}) \Big{]}.
\label{eigenvalue_equation_jump2}
\end{split}
\end{equation}
Here, for any potential function $V(\bv a)$, we define a modified
transition rate $u_h^{V}(\bv a\rightarrow \bv a^{\prime})$ as
\begin{equation}
\begin{split}
& u^{V}_h(\bv a\rightarrow \bv a^{\prime}) \\
& \equiv w(\bv a \rightarrow \bv a^{\prime})e^{-V(\bv a^{\prime})/2
 + V(\bv a)/2+ h\alpha(\bv a,\bv a^{\prime})}.
\end{split}
\label{jumpmodifiedtransition}
\end{equation}
The corresponding escape rate $\lambda^{V}_h$ becomes
\begin{equation}
\lambda^{V}_h(\bv a)
=\sum_{\bv a^{\prime}} u^{V}_h(\bv a\rightarrow \bv a^{\prime}),
\label{jumpmodifiedescaperate}
\end{equation}
and (\ref{eigenvalue_equation_jump2}) is rewritten as
\begin{equation}
G(h)=\lambda^{V_{\rm max}}_h(\bv a) - \lambda(\bv a).
\label{jump1}
\end{equation}
This expression might be known,
although we do not find its explicit presentation
in previous studies. 


Now, our  main claim, which is  far from trivial, is the 
variational expression
\begin{equation}
\begin{split}
&G(h)  = \max_{V} \left \langle \lambda^{V}_h - 
 \lambda \right \rangle _{\rm st}^{u^{V }_h},
 \label{jump2}
\end{split}
\end{equation}
and the relation
\begin{equation}
\frac{\partial G(h)}{\partial h}
= \sum_{\bv a,\bv a^{\prime}}  P^{u^{V_{\rm opt}}_h}_{\rm st}(\bv a)u^{V _{\rm opt}}_h(\bv a\rightarrow \bv a^{\prime})
\alpha(\bv a,\bv a^{\prime})
\label{jump4}
\end{equation}
with
\begin{equation}
V_{\rm opt} 
= \underset{V}{\rm argmax} \left \langle \lambda^{V}_h - 
\lambda \right \rangle _{\rm st}^{u^{V }_h}.
\end{equation}


The derivation is the following.
First, we consider the variation $V\rightarrow V+\delta V$ in
$\left \langle \lambda^{V}_h - 
 \lambda \right \rangle _{\rm st}^{u^{V }_h}$,
 which is expressed as
\begin{equation}
\begin{split}
&\delta \left [ \left \langle \lambda^{V}_h - 
 \lambda \right \rangle _{\rm st}^{u^{V }_h} \right ]\\
& =\sum_{\bv a} \delta P_{\rm st}^{u_h^V}(\bv a)
 \left [ \lambda^{V}_h(\bv a) - \lambda(\bv a) \right ]
 +\sum_{\bv a} P_{\rm st}^{u_h^V}(\bv a) \delta \lambda^{V}_h(\bv a). 
\label{jumpvariation}
\end{split}
\end{equation} 
From  (\ref{jumpmodifiedtransition}) and (\ref{jumpmodifiedescaperate}),
we rewrite the second term of (\ref{jumpvariation}) as
\begin{equation}
-\frac{1}{2} \sum_{\bv a} \sum_{\bv a^{\prime}} P_{\rm st}^{u_h^V}(\bv a) u_{h}^V(\bv a
 \rightarrow \bv a^{\prime}) \left [\delta V(\bv a^{\prime}) - \delta V(\bv a) \right ].
\end{equation} 
Since this becomes zero from
the condition of the stationary distribution,
the variational equation
$\delta \left [ \left \langle \lambda^{V}_h - 
 \lambda \right \rangle _{\rm st}^{u^{V }_h} \right ]=0$ leads to
\begin{equation}
K=\lambda^{V}_h(\bv a) - \lambda(\bv a),
\label{jump3}
\end{equation}
where $K$ is a constant corresponding to a Lagrange multiplier.
Here,  (\ref{jump3}) is equivalent to 
\begin{equation}
\sum_{\bv a^{\prime}} e^{-V_{\rm opt} (\bv a^{\prime})/2} \mathcal A_h( \bv a^{\prime},\bv a)
=K e^{-V_{\rm opt}(\bv a)/2}.
\end{equation}
By comparing this with (\ref{eigenvalue_equation_jump}),
we find that $V_{\rm opt}=V_{\rm max}$ as the result of the 
Perron-Frobenius theory. 

Next, we consider the modified system with the transition rate
$u^{V _{\rm opt}}_h(\bv a\rightarrow \bv a^{\prime})$.
By multiplying the both sides of (\ref{jump1}) by 
$P^{u^{V _{\rm opt}}_h}_{\rm st}(\bv a)$ and summing over $\bv a$,
we obtain
\begin{equation}
G(h)=\left \langle \lambda^{V_{\rm opt}}_h - 
\lambda \right \rangle _{\rm st}^{u^{V _{\rm opt}}_h},
\label{G(h)_max}
\end{equation}
where we have used $V_{\rm opt}=V_{\rm max}$. We thus conclude 
(\ref{jump2}). Furthermore, by differentiating (\ref{jump1}) 
with respect to $h$, multiplying the  both  sides by 
$P^{u^{V _{\rm opt}}_h}_{\rm st}(\bv a)$, and summing over $\bv a$, 
we derive 
\begin{equation}
\begin{split}
&\frac{\partial G(h)}{\partial h} = \sum_{\bv a,\bv a^{\prime}}  P^{u^{V_{\rm opt}}_h}_{\rm st}(\bv a)u^{V _{\rm opt}}_h(\bv a\rightarrow \bv a^{\prime})
\alpha(\bv a,\bv a^{\prime})\\
&-\frac{1}{2}\sum_{\bv a,\bv a^{\prime}}  P^{u^{V_{\rm opt}}_h}_{\rm st}(\bv a)u^{V _{\rm opt}}_h(\bv a\rightarrow \bv a^{\prime})
\left (\frac{\partial V_{\rm opt}(\bv a^{\prime})}{\partial h}- \frac{\partial V_{\rm opt}(\bv a)}
{\partial h}\right ).
\label{jump_diff1}
\end{split}
\end{equation}
The second term of (\ref{jump_diff1}) turns out to be zero from
the condition of the stationary distribution.
The final expression is (\ref{jump4}).

At the end of this Appendix, we point out that the expression (\ref{jump2})
is similar to the formula of the large deviation function
of the empirical measure in Markov jump processes \cite{Maes_Linear_Response}.
This similarity corresponds to the relation between (\ref{variational_expression3}) 
and (\ref{relation_DV3}) in Langevin equations.


\end{document}